\documentclass[12pt, preprint]{aastex}

\usepackage{subfigure}

\shorttitle{Parallaxes from CTIOPI}
\shortauthors{Jao et al.}

\begin{document}

\title{The Solar Neighborhood XLII. Parallax Results from the CTIOPI
  0.9-m Program --- Identifying New Nearby Subdwarfs Using Tangential
  Velocities and Locations on the H-R Diagram}

\author{Wei-Chun Jao\altaffilmark{1}}
\affil{Department of Physics and Astronomy, Georgia State University, Atlanta, GA 30302}
\email{jao@astro.gsu.edu}

\author{Todd J. Henry\altaffilmark{1}}
\affil{RECONS Institute, Chambersburg, PA 17201}
\email{toddhenry28@gmail.com}

\author{Jennifer G. Winters\altaffilmark{1}}
\affil{Harvard-Smithsonian Center for Astrophysics, Cambridge, MA 02138}
\email{jennifer.winters@cfa.harvard.edu}

\author{John P. Subasavage\altaffilmark{1}}
\affil{United States Naval Observatory, Flagstaff Station, Flagstaff, AZ 86601}
\email{jsubasavage@nofs.navy.mil}  

\author{Adric R. Riedel\altaffilmark{1}}
\affil{Astronomy Department, California Institute of Technology, Pasadena, CA 91125}
\email{arr@astro.caltech.edu}

\author{Michele Silverstein\altaffilmark{1}}
\affil{Department of Physics and Astronomy, Georgia State University,  Atlanta, GA 30302}
\email{silverstein@astro.gsu.edu}

\and 

\author{Philip A. Ianna\altaffilmark{1}}
\affil{RECONS Institute, Chambersburg, PA 17201}
\email{philianna3@gmail.com}

\altaffiltext{1}{Visiting Astronomer, Cerro Tololo Inter-American
  Observatory.  CTIO is operated by AURA, Inc.\ under contract to the
  National Science Foundation.}

\begin{abstract}

Parallaxes, proper motions, and optical photometry are presented for
51 systems made up 37 cool subdwarf and 14 additional high proper
motion systems.  Thirty-seven systems have parallaxes reported for the
first time, 15 of which have proper motions of at least 1$\arcsec$
yr$^{-1}$.  The sample includes 22 newly identified cool subdwarfs
within 100 pc, of which three are within 25 pc, and an additional five
subdwarfs from 100-160 pc.  Two systems --- LSR 1610-0040 AB and LHS
440 AB --- are close binaries exhibiting clear astrometric
perturbations that will ultimately provide important masses for cool
subdwarfs.

We use the accurate parallaxes and proper motions provided here,
combined with additional data from our program and others to determine
that {\it effectively all nearby stars with tangential velocities
  greater than 200 km s$^{-1}$ are subdwarfs}.  We compare a sample of
167 confirmed cool subdwarfs to nearby main sequence dwarfs and
Pleiades members on an observational Hertzsprung-Russell diagram using
$M_V$ vs.~$(V-K_{s})$ to map trends of age and metallicity.  We find
that subdwarfs are clearly separated for spectral types K5--M5,
indicating that the low metallicities of subdwarfs set them apart in
the H-R diagram for $(V-K_{s})$ = 3--6.  We then apply the tangential
velocity cutoff and the subdwarf region of the H-R diagram to stars
with parallaxes from {\it Gaia} Data Release 1 and the MEarth Project
to identify a total of 29 new nearby subdwarf candidates that fall
clearly below the main sequence.

\end{abstract}

\keywords{astrometry --- solar neighborhood --- stars: distances ---
  stars: late-type --- subdwarfs}

\section{Introduction}

Cool subdwarfs of spectral types G, K, and M are Galactic relics with
relatively low metallicities compared to their dwarf counterparts
\citep{Chamberlain1951, Mould1976}.  Unlike the abundant metal-rich
dwarfs in the solar neighborhood, there are currently only three
confirmed subdwarf systems within 10 pc: $\mu$ Cas AB, CF UMa, and
Kapteyn's Star \citep{Monteiro2006}, making them minorities in our
solar neighborhood.  Because of their scarcity and intrinsic
faintness, fewer key stellar parameters, such as radius and mass, have
been measured for subdwarfs compared to the dwarfs.  For example,
\cite{Jao2016} showed that there are only five nearby confirmed
subdwarf binaries with measured dynamical masses, compared to at least
five times that for M dwarf dynamical masses alone \citep{Henry1999,
  Benedict2016}.  Direct measurements of stellar radii are almost
entirely for main sequence dwarfs \citep{Segransan2003, Berger2006,
  Lopez2007, Torres2010, Boyajian2012}.  The $\mu$ Cas A is the only
subdwarf\footnote{A compendium of eclipsing binaries by
  \cite{Lopez2007} and \cite{Torres2010} shows a few stars with
       [Fe/H]$\le-$0.5, but almost all of them are early type
       subdwarfs. Of particular note, GJ 630.1 AB (CM Dra AB) is an
       eclipsing binary with [Fe/H]=$-$0.67 \citep{Lopez2007}, but
       \cite{Hawley1996} assigned it spectroscopically as a M4.5 V,
       i.e, a main sequence star. Furthermore, a wide third component
       is a white dwarf with an estimated age of 3 Gyrs
       \citep{Bergeron1997}, so it is not old enough to be a subdwarf.
       This shows inconsistent results between metallicity, spectral
       classification, and ages. Hence, we do not consider GJ 630.1 AB
       to be a subdwarf system. }  with interferometric measurement of
its radius \citep{Boyajian2008}, but it is a G-type subdwarf. Thus, to
understand the nature of the metal-poor stars that formed early in the
history of the Galaxy, it is important to identify more nearby
subdwarfs so that the most basic stellar parameters of masses and
radii can be determined.

In order to reveal nearby subdwarfs, the Cerro Tololo Inter-american
Observatory Parallax Investigation (CTIOPI) carried out by RECONS
(REsearch Consortium On Nearby Stars)\footnote{\url{www.recons.org}} has
targeted subdwarf candidates with high proper motions extracted from
various catalogs and surveys \citep{LHS,Giclas1971, Giclas1979,
  Pokorny2003, Hambly2004, Scholz2004b, Lepine2005, Deacon2005,
  Gizis2011}.  Here we present the first parallaxes for 37 stellar
systems selected from these surveys and revised parallaxes for 14
additional systems.  As in previous papers in {\it The Solar
  Neighborhood} series, the overlap in the samples of fast-moving and
low-metallicity stars makes it natural to combine the two types of
objects in this paper with the primary goal to unveil more nearby
missing subdwarfs.

\section{Observations and Data Reduction}

We used the CTIO/SMARTS 0.9-m to measure parallaxes and optical
photometry in the $VRI$ filters.  The telescope has a 2048$\times$2046
Tektronix CCD camera with 0\farcs401 pixel$^{-1}$ plate scale
\citep{Jao2005}.  For both astrometric and photometric observations,
we used the central quarter of the chip, yielding a 6\farcm8 square
field of view.  We used the Johnson $V$ and Kron-Cousins $RI$ filters
for parallax measurements to maximize the number of suitable reference
stars in the field; because of their relative faintness, 31 of the 51
systems were observed in the $I$ band.  For the 51 systems discussed
here, astrometric series spanned 2--15 years with a median of 5 years.
We also obtained $VRI$ photometry of the targets and parallax
reference stars through the same filters.  The photometry was used to
characterize the stars, remove differential color refraction offsets
in the astrometry, and correct the relative parallaxes to absolute
parallaxes via photometric distance estimates of the reference stars.

Bias and dome flat frames were taken nightly for basic image
reductions and calibrations.  Details of our observing methodology and
data reduction, including astrometric, photometric, and spectroscopic
techniques, are discussed in previous parallax papers of {\it The
  Solar Neighborhood} series; in particular, see \citep{Jao2005} for
astrometry protocols and \citep{Winters2015} for photometry methods.

\section{Results}
\label{sec:results}

\subsection{Astrometry Results}
\label{sec:astro}

The astrometry results are presented in Table~\ref{tbl:pi.result},
where we provide details about the astrometric observations.  The
first column gives the target identifiers, followed by coordinates
(column 2), filters used (3), number of seasons observed (4), number
of frames used in reductions (5), time coverage (6), the total time
spans (7), the number of reference stars (8), relative parallaxes (9),
parallax corrections (10), absolute parallaxes (11), proper motions
(12), position angles of the proper motions (13), and the derived
tangential velocities (14).  An exclamation point in the Note column
indicates that additional details about that system are provided in
Section~\ref{sec:notes}.

High proper motion stars fall into two astrophysically interesting
categories --- nearby stars and those with intrinsically high space
velocities, typically subdwarfs.  Among the 37 systems for which we
provide the first parallaxes here, 15 are moving faster than
1$\arcsec$ yr$^{-1}$, including seven subdwarfs, seven main sequence
red dwarfs, and a brown dwarf 2MA 1506+1321.

Parallax errors are less than 2 mas for all but five systems.  Among
the 37 systems with parallaxes reported for the first time here, nine
are within 25 parsecs (pc) and nine more are between 25 and 60 pc.
The latter horizon is being used to build a volume-complete sample of
the nearest cool subdwarfs, and includes nine new subdwarfs first
identified to be within 60 pc here --- LEHPM 1-4592, LHS 1257, LHS
1490, LHS 2096, LHS 2099, LHS 2140, LHS 2904, LSR 0609+2319, and SSS
1358-3938.  The remaining 19 systems are between 60 and 160 pc.
Overall, the RECONS astrometry program on both the 0.9-m and 1.5-m at
CTIO have added 26 new cool subdwarfs within 60 pc \citep{Costa2005,
  Jao2005, Jao2011} since 1999, including this work. This constitutes
a significant increase of 25\% to the previously known sample
\citep{Hipparcos, YPC, Burgasser2008, Schilbach2009,
  Smart2010}\footnote{A comprehensive discussion of the entire 60 pc
  cool subdwarf sample is planned for a future paper in this series.}.

\subsection{Photometry Results}
\label{sec:phot}

Results of our $VRI$ photometry as well as the near-IR photometry from
the Two Micron All Sky Survey \citep{2mass} are presented in
Table~\ref{tbl:phot.result}.  The first two columns provide
identifiers, followed by the $VRI$ magnitudes (columns 3,4,5), the
number of $VRI$ observations (6), the filter in which parallax frames
were taken (7), the variability in that filter (8), the $JHK_{s}$
photometry (9,10,11), the spectral type (12), and the spectral type
reference (13).

Stars were observed in $VRI$ filters, spanning magnitude ranges of $V$
= 11.49--20.25, $R$ = 10.49--19.30, and $I$ = 9.31--18.34.  All stars
were observed in all three filters except 2MA1506+1321, which is too
faint in $V$ to be observed effectively at the 0.9-m telescope, but
for which $R$ and $I$ magnitudes are provided.  All stars except SIP
1540-2613 were observed 2--4 times.  As described in detail in
\cite{Winters2011}, the mean standard deviations of our multi-epoch
photometry are typically $\sim$0.03 mag in $V$ and $\sim$0.02 mag in
$R$ and $I$ bands.  This is true regardless of magnitude, as fainter
stars are simply observed with longer integrations to increase
signal-to-noise.

The combination of our astrometry and photometry results allows us to
place the sample of stars on the observational H-R diagram shown in
Figure~\ref{fig:HRdiagram}, which uses $M_V$ and $(V-K_S)$.  Stars
within 25 pc are represented with gray points, overlaid with the
sample stars in black.  Several noteworthy stars discussed in
Section~\ref{sec:notes} are circled and labeled in red.

\subsection{Variability Results}
\label{sec:varia}

The long-term data series of images taken for astrometry of the
observed stars permits an evaluation of their photometric variability
in the filter used for the observations.  Listed in column 8 of
Table~\ref{tbl:phot.result} are the variability results for each
target.

\cite{Jao2011} first reported the long term variability of our
parallax stars with coverage from 2--10 years and found that the 22
cool subdwarfs investigated at the time were substantially less
photometrically variable than the 108 main sequence red dwarf
examined.  \cite{Hosey2015} then expanded the variability study to 264
M dwarfs and found that only 8\% of M dwarfs are photometrically
variable by at least 20 mmag.  Details of the data reduction processes
used to determine variability can be found in those two papers.  The
median variability of the 42 subdwarfs in this work is only 9 mmag,
with only one subdwarf, LHS2852 (24 mmag), having a variability
greater than 20 mmag.  Hence, we reconfirm the conclusion we made in
\cite{Jao2011} that subdwarfs are, in general, photometrically quiet.
We note, however, that because of the faintness of the targets in this
sample, most (26 of 42) of the subdwarfs were observed in the $I$ band
for parallax observations, so the variability for those objects is
likely to be lower than stars observed in the $V$ band, which includes
potentially variable H$\alpha$ emission.

\subsection{Spectroscopy of Cool Subdwarfs}
\label{sec:spect}

Spectral types from the literature, including many of our own results,
are given in columns 12 and 13 of Table~\ref{tbl:phot.result}.
Although we do not present any new spectra in this paper, for stars
with no spectra available we can make informed estimates of luminosity
classes based on the astrometry and photometry data presented here, as
discussed in Section~\ref{sec:astro} and~\ref{sec:phot}.  We assign
luminosity class of ``VI'' to 14 stars we now identify to be
subdwarfs, and three as main sequence stars of class ``V''.  As
discussed in \cite{Jao2008}, the ``sd'' prefix often used to classify
cool subdwarfs is the same prefix used for hot subdwarfs, even though
they are completely different types of stellar objects.  The mixed use
of ``sd'' is unique in spectral classification, so we prefer the VI
designation.  In support of this spectral type moniker,
Figures~\ref{fig:HRdiagram},\ref{fig:newHR}, \ref{fig:GAIA},
and~\ref{fig:MEarth} all clearly show a different luminosity class on
the Hertzsprung–Russell (H-R) diagram for a given $V-K_{s}$ between
(at least) 3 and 6.

\section{Notes on Individual Systems}
\label{sec:notes}

Here we provide additional details of systems worthy of note, listed
in order of RA.

{\bf 0342+1231 (LHS 178)} The Yale Parallax Catalog (YPC,
\citealt{YPC}) provides a parallax of 45.1$\pm$12.0 mas for this star
and we find 40.19$\pm$2.49 mas, resulting in a weighted mean value of
40.39$\pm$2.44 mas.

{\bf 0432-3947 (LHS 1678)} We detect a possible perturbation with a
period of a few years in the astrometric series spanning 12 years, but
because it is slight, we have not removed the perturbation to
calculate the parallax presented here.  Until there is further
evidence to support the existence of a currently unseen companion, we
consider this is a single star.


{\bf 0559+0410 (G 99-48AB)} \cite{Goldberg2002} found this system to
be a double-lined spectroscopic binary, and \cite{Soubiran2010}
determined it have [Fe/H]=$-$1.80.  Our parallax of 8.06$\pm$1.85 mas
is consistent with that provided in {\it Gaia} Data Release 1
(hereafter DR1), 7.06$\pm$0.25 mas.

\begin{figure}
\centering
\includegraphics[angle=-90, scale=0.6]{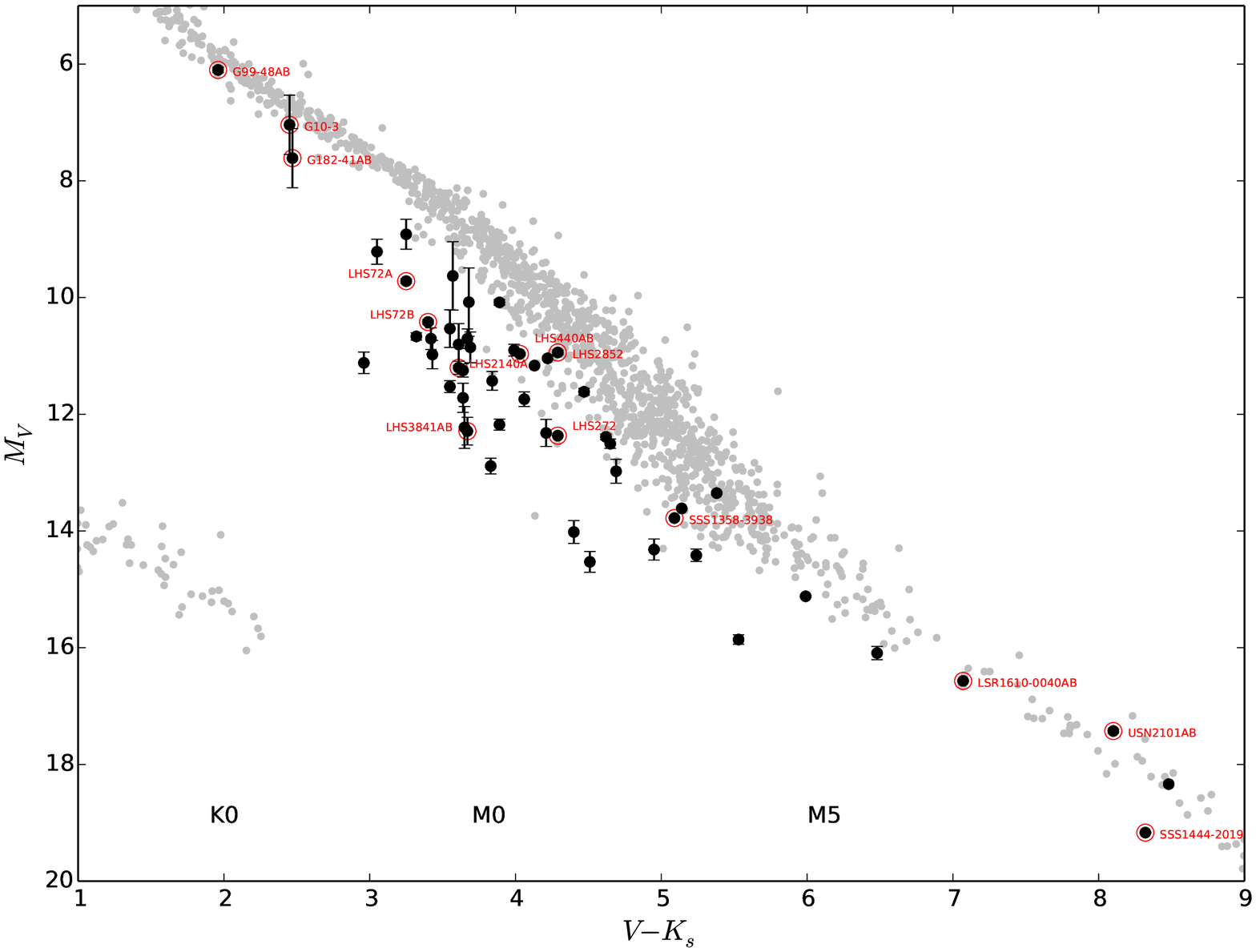}

  \caption{An observational H-R diagram, using $M_{V}$ vs.~$V-K_{s}$,
    is shown for 53 stars (dark filled circles) in the 51 systems
    outlined in Tables~\ref{tbl:pi.result} and~\ref{tbl:phot.result}.
    Two objects, the brown dwarf 2MA 1506$+$1321 and the white dwarf
    LHS 2139, do not have both $V$ and $K_s$ magnitudes and are not
    shown.  For comparison, gray points represent stars within 25 pc,
    with data taken primarily from the Yale Parallax Catalog, {\it
      Hipparcos} results, our RECONS astrometry/photometry program,
    the MEarth Project, and {\it Gaia} DR1.  It is clear that most of
    the stars presented in this work are below the main sequence and
    are subdwarfs.  Red circles highlight interesting systems
    discussed in Section~\ref{sec:notes}.  Errorbars in the horizontal
    direction are smaller than the points.}

\label{fig:HRdiagram}
\end{figure}

{\bf 0905-2201 (LHS 2099/2100)} This pair of subdwarfs is separated by
6\farcs6 at a position angle of 100$^{\circ}$, corresponding to 326 AU
at a distance of 49.4 pc for the weighed mean parallax of 20.24 $\pm$
0.78 mas.  The secondary has $M_V$ = 15.69 and $(V-K_{s})$ = 5.53,
placing it well below the main sequence in Figure~\ref{fig:HRdiagram}
and making it one of the reddest subdwarfs in the sample.

{\bf 0925+0018 (LHS 2140 and LHS 2139)} \cite{Gizis1997} first
reported this common proper motion binary to consist of a subdwarf
primary (LHS 2140) and a white dwarf secondary (LHS 2139).  LHS 2139
is too faint in our images to measure a reliable parallax, so we adopt
the parallax of LHS 2140 for both components.  This is one of the very
few known subdwarf$+$white dwarf binaries with parallaxes in the solar
neighborhood \citep{Monteiro2006}.

{\bf 0943-1747 (LHS 272)} The updated parallax (69.4$\pm$1.02 mas) has
a longer time coverage than previously reported in \cite{Jao2011}, and
supercedes our previous result.  This is the fourth nearest known
subdwarf system of any spectral type, ranking behind Kapteyn's Star
(GJ 191, M type), $\mu$ Cas AB (GJ 53AB, G and M types), and GJ 451 (K
type).  

{\bf 1005-6721 (WT 248)} \cite{Faherty2012} reported a parallax of
30.6$\pm$4.6 mas for this object.  Our parallax of 41.64$\pm$2.23
places the system within 25 pc, but the weighted mean of
39.54$\pm$2.00 mas is still slightly less than 40 mas.

{\bf 1110-0247 (G 10-3)} \cite{Bidelman1985} classified this star as a
K2 dwarf, but it undoubtedly a subdwarf given that \cite{Latham2002}
report the star to have $[m/H]=-$2.0.  YPC provides a parallax of
28.1$\pm$13.4 mas and we find 7.48$\pm$1.86, yielding a weighted mean
value of 7.87$\pm$1.84 mas.

{\bf 1234+2037 (LHS 334)} Our parallax of 17.48$\pm$1.99 mas is
consistent with that of \cite{Smart2010}, who reported a parallax of
22.1$\pm$3.9 mas.  The new weighted mean parallax is 18.43$\pm$1.77
mas.

{\bf 1358-3938 (SSS 1358-3938)} This star has a proper motion of
nearly 2$\arcsec$ yr$^{-1}$ and we provide the first parallax here,
88.66$\pm$0.80 mas.  The star is located on the edge of the main
sequence band in the H-R diagram of Figure~\ref{fig:HRdiagram}.  The
photometric distance determined using $VRIJHK$ and the relations of
\cite{Henry2004} place this star at 21.2 pc, but our parallax puts it
at 11.3 pc.  With no previous spectral type available, this large
distance mismatch implies this star is likely a subdwarf and we assign
it as ``VI''.  If spectroscopically confirmed, this star would replace
LHS 272 as the fourth closest subdwarf system and the third closest M
type subdwarf.

{\bf 1402-2431 (LHS 2852)} \cite{Gizis1997b} identified this star to
be a subdwarf, but it is located on lower edge of the main sequence in
the H-R diagram of Figure~\ref{fig:HRdiagram}.  \cite{Haakonsen2009}
found it to have ROSAT X-ray detection of 0.15$\pm$0.03 cnt s$^{-1}$.
In comparison, the known young star AP Col \cite{Riedel2011} has
0.43$\pm$0.05.  In addition, LHS2852 varies by 24 mmag in the $R$
band, the largest variability seen among the 42 subdwarfs studied
here.  It would be unusual for a subdwarf to be so variable and
detected in X-rays, so followup spectroscopy is needed to confirm
whether or not the star is a subdwarf.  YPC provides a parallax of
39.4$\pm$19.9 mas and we find 57.99$\pm$1.88 mas, yielding a weighted
mean value of 57.83$\pm$1.87 mas.

{\bf 1444-2019 (SSS 1444-2019)} This high proper motion (nearly 3.5
$\arcsec$ yr$^{-1}$) star was first detected and classified as an M9
subdwarf by \cite{Scholz2004}.  Recently, \cite{Kirkpatrick2016}
re-classified it as a L0 subdwarf.  Our parallax of 60.18$\pm$1.62 mas
is consistent with the parallaxes reported by \cite{Schilbach2009}
(61.67$\pm$2.12 mas) and \cite{Faherty2012} (61.2$\pm$5.1 mas).
Together, the three values result in a weighted mean parallax of
60.76$\pm$1.25 mas.

{\bf 1455-1533 (LHS 385)} YPC provides a parallax of 20.4$\pm$5.8 mas
and we find 24.43$\pm$1.51 mas, resulting in a weighted mean value of
24.17$\pm$1.46 mas.

{\bf 1506+1321 (2MA1506+1321)} \cite{Gizis2000} reported this object
to be an L3 dwarf, implying that it is a brown dwarf because it is
later than the L2 type found by \cite{Dieterich2014} at the
stellar/substellar boundary.  It is too faint for $V$ band photometry
at the 0.9-m, so no DCR correction was made for this field; however,
because this field was observed in the $I$ band, any DCR correction is
minimal.  \cite{Gagne2014} flagged this object with a 100\%
probability to be a young field object.  The object shows three signs
of youth: (1) a triangular-shaped H-band continuum, (2)
redder-than-normal colors for its assigned spectral type, and (3)
signs of low gravity from atmospheric model fitting.  With a
relatively slow tangential velocity of 58.3 km s$^{-1}$, it does not
have the typical high velocity of an old subdwarf, so it is likely a
young brown dwarf for which we provide the first parallax of
87.08$\pm$1.58 mas.

{\bf 1539-5509 (LHS 401)} YPC provides a parallax of 38.4$\pm$9.6 mas
and we find 38.67$\pm$1.68 mas, resulting in a weighted mean value of
38.66$\pm$1.65 mas.

\begin{figure}
\centering 
 \subfigure[]{\includegraphics[scale=0.35,angle=90]{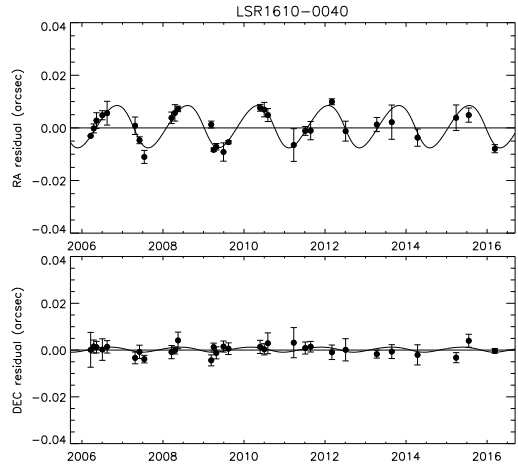}
  \label{fig:lsr1610}} 
\subfigure[]{ 
  \includegraphics[scale=0.35,angle=90]{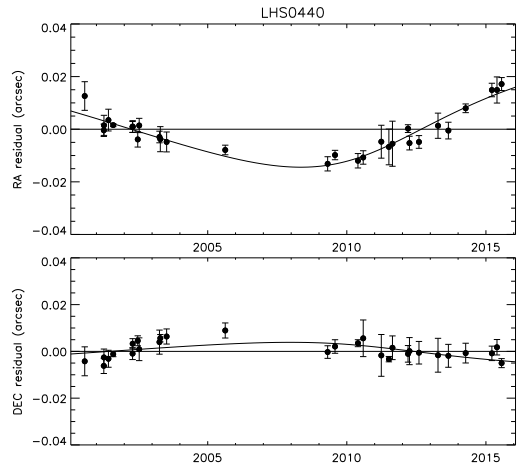}
  \label{fig:lhs0440}} 

  \caption{Nightly mean astrometric residuals in right ascension
  and declination are shown for LSR 1610-0040 AB and LHS 440 AB.
  The astrometric signatures of each system's proper motion and
  parallax have been removed.  Solid-line curves show the best fits
  to the perturbations and these fits have been removed when
  determining the proper motions and trigonometric parallaxes
  presented here. [To Editor: Please arrange two plots
  horizontally.]}

\label{fig:perturb}
\end{figure}

{\bf 1610-0040 (LSR1610-0040 AB)} is an important subdwarf system,
given that it promises to yield accurate masses for a pair of very
cool subdwarfs.  This system was first detected and classified as an
early type L subdwarf by \cite{Lepine2003a}.  \cite{Cushing2006} later
showed the system to have a peculiar spectrum with an ambiguous
assignment of dwarf or subdwarf type.  \cite{Dahn2008} reported the
first trigonometric parallax of 31.02$\pm$0.26 mas, detected a clear
photocentric orbit with a semi-major axis of 8.91 mas, and classified
it as type ``sd?M6pec''.  Recently, \cite{Koren2016} estimated the
masses of the primary and unseen secondary using updated astrometry
from \cite{Dahn2008} and radial velocity data from \cite{Blake2010}
and found masses of 0.09--0.10 $M_{\odot}$ and 0.06 -- 0.075
$M_{\odot}$.  Our dataset of 9.98 years also detects a large
astrometric perturbation (see Figure~\ref{fig:lsr1610}) with a period
of 634.8 days, nearly identical to the period of 633 days found by
\cite{Koren2016}, even though their timespan is 10.2 years. We measure
a photocentric semi-major axis of 8.25$\pm$0.63 mas, but
\cite{Koren2016} has 9.89$\pm$0.25 mas. Although both USNO and CTIO
parallax programs use ``I'' filters, their bandpasses are
different. The effective central wavelength and filter band width for
USNO and CTIO are 8074\AA/1890\AA and 8118\AA/1415\AA,
respectively. These two different bandpasses may cause the slight
difference in the photocentric semi-major axis.

\cite{Schilbach2009}
reported a parallax of 33.1$\pm$1.32 mas, and an updated USNO parallax
of 30.73$\pm$0.34 mas, calculated using an Markov chain Monte Carlo
simulation, is given in \cite{Koren2016}.  After removing the
perturbation of the photocenter shown in Figure~\ref{fig:lsr1610}, we
find a parallax of 32.26$\pm$0.48 mas.  The final weighted mean
parallax from these three measurements is 31.32$\pm$0.27 mas.

{\bf 1718-4326 (LHS 440 AB)} \cite{Jao2008} reported this to be a M1.0
subdwarf and \cite{Jao2011} first reported the possible unseen
companion based upon a photocentric perturbation.  We have extended
the coverage from 9 years in \cite{Jao2011} to $\sim$15 years in this
work, yielding the perturbation curve shown in
Figure~\ref{fig:lhs0440}.  Because of the very long orbital period for
the system, we do not detect a full photocentric orbit yet.  The
parallax of 39.56$\pm$1.02 mas presented in Table~\ref{tbl:pi.result}
has had the perturbation removed and supersedes previously reported
values in \cite{Jao2005} and \cite{Jao2011}.

\begin{figure}
\centering
\includegraphics[scale=0.6]{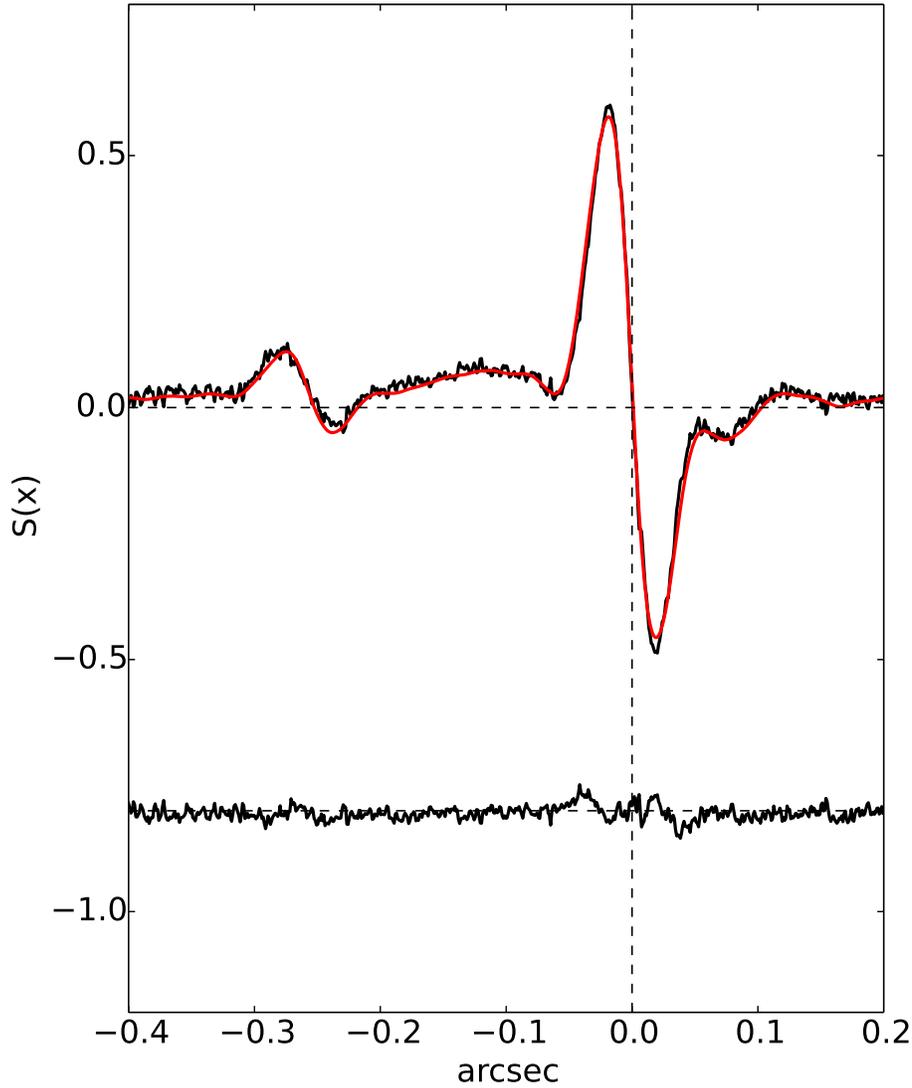}

  \caption{The ``S-curve'' along the X-axis from a {\it Hubble
  Space Telescope} Fine Guidance Sensor observation of LHS 440 AB.
  The black line on the top represents the data and the red line is
  the best fit.  The residuals to the fit are shown on the bottom
  of the plot.  The secondary is clearly seen at a separation of
  256 mas with a $\Delta_{F583W}=$2.03.  Note that the X-axis is
  {\it not} the direction of RA because of the {\it HST's} roll
  angle at the observation epoch.  The binary is not resolved along
  the Y-axis, so is not shown here.}

\label{fig:fgs}
\end{figure}

We used the FGS1r (Fine Guidance Sensors) on the {\it Hubble Space
  Telescope} to resolve this system in Cycle 16 on April 10th, 2009,
with scan duration of 1301 sec and scan lengths of 6\arcsec.
Observations were made through the F583W filter, which provides
magnitude differences similar to the $V$ band in the Johnson system.
The standard {\it strfits} routine in the IRAF/STSDAS package was used
to measure the separation, position angle, and magnitude difference
between the two components.  Additional details of the reduction
procedure can be found in the {\it HST/FGS Data Handbook}
\url{http://www.stsci.edu/hst/fgs/documents/datahandbook/}.  The
calibrator is LHS 73, which is also a subdwarf and was observed in the
same HST observing Cycle.  LHS 440 AB was successfully resolved along
the X-axis, as shown in Figure~\ref{fig:fgs}.  The pair was not
resolved along the Y-axis at this epoch, indicating a separation of
$\pm$ 10 mas, so we cannot calculate the companion's separation and
position angle.  The magnitude difference is 2.03 mag in the F583W
filter.  \cite{Henry1999} presented a conversion from $\Delta
m_{F583W}$ to $\Delta V$ using $B-V$ colors, but no $B$ photometry is
available for the components, and the relations used in the absence of
$B$ photometry apply to dwarfs rather than subdwarfs.  Assuming for
now that $\Delta V$ $\approx$ 2, we find $M_V$ = 11.13 and 13.13 for
the components.  From the mass-luminosity relation for main sequence
red dwarfs of \cite{Benedict2016}, this implies masses of 0.33 and
0.19 $M_{\odot}$, but again because these are subdwarfs rather than
dwarfs, we emphasize that these should be considered only crude
estimates.

{\bf 1750-5636 (LHS 456)} YPC provides a parallax of 39.2$\pm$12.6 mas
and we find 39.87$\pm$1.43, resulting in a weighted mean value of
39.86$\pm$1.42 mas.

{\bf 1809-6154 (SCR1809-6154 AB)} The separation of these two stars is
19\farcs2 at a position angle of 270.6$^{\circ}$.  Because the primary
star is very close to a background star, no variability is reported
for the primary and only frames with seeing less than 1\farcs4 were
kept for the astrometric reduction.  The weighted mean parallax for
the two components is 5.50$\pm$1.16 mas.

{\bf 1809+2755 (G 182-41AB)} This is a double-lined spectroscopic
binary with [Fe/H]=$-$1.0 \citep{Goldberg2002}, clearly indicating
that it is a subdwarf.  Our parallax of 9.96$\pm$2.33 mas places the
pair at $\sim$100 pc, indicating that the resolution needed to
determine masses will prove difficult.

{\bf 2101+0307 (USN 2101+0307AB)} The combined photometry make this
system elevated on the H-R diagram. We do not have a spectrum for this
system. Becasue of its location on the H-R diagram, we temporarily
assign its luminosity class as ``V''.

{\bf 2239-3615 (LHS 3841AB)} \cite{Friedrich2000} showed this star to
have a combined spectrum of helium-rich white dwarf and an M dwarf
using a spectrograph with a coverage of 3800\AA~to 9200\AA.
\cite{Reid2005} later identified the red dwarf to be an M2.5 subdwarf
with a wavelength coverage of 6200\AA~to 7500\AA~and estimated the
distance to be 19 pc.  However, we determine a parallax of
12.13$\pm$1.47 mas, placing the system at $\sim$80 pc.  The erroneous
distance estimate was likely due to excess flux from the white dwarf
not being considered.  \cite{Farihi2010} used the Advanced Camera for
Surveys High-Resolution Camera on the {\it Hubble Space Telescope} in
an attempt to resolve this system using the F814W filter, but LHS3841
AB was not resolved.  Over 4.2 years of astrometric observations, we
do not detect any perturbation of the photocenter position.

{\bf 2343-2409 (LHS 72 and LHS 73)} Both components are subdwarfs
\citep{Rodgers1974, Reyle2006, Jao2008}, separated by 1\farcm6 at
position angle of 153$^{\circ}$.  The primary, LHS 72, has a parallax
of 37.6$\pm$8.9 mas in the YPC, but no separate parallax is given for
LHS 73.  We find parallaxes of 35.20$\pm$1.63 mas and 32.73$\pm$1.44
mas for the primary and secondary, respectively, which agree within
2$\sigma$.  The weighted mean of all three parallax measurements is
33.87$\pm$1.07 mas.

\section{Discussion: Observational Differences between Dwarfs and Subdwarfs}
\label{discussion}

Subdwarfs are low-metallicity stars that have historically been
discovered through proper motion surveys, spectroscopic surveys, or
color index searches.  Historically, the extensive proper motion
surveys by Luyten and Giclas have been the primary sources for finding
subdwarfs \citep{Ryan1991, Bessell1982, Gizis1997b}.  Recent new
proper motion surveys with fainter magnitude limits like SuperBLINK
\citep{Lepine2005}, SuperCOSMOS-RECONS \citep{Subasavage2005}, and
SIPS \citep{Deacon2005} have discovered many new metal-deficient high
proper motion subdwarf candidates.  To confirm that these candidates
are, indeed, cool subdwarfs, spectroscopic observations are typically
necessary, such as those by \citep{Scholz2004, Lepine2007, Jao2008,
  Zhang2013}.  Large sky spectroscopic surveys like SDSS and LAMOST
have also provided systematic ways to identify cool subdwarfs
\citep{West2004, Zhong2015}.  Finally, because subdwarfs have
different colors from dwarfs, infrared colors from the all sky
infrared surveys 2MASS and WISE have been used to identify local
stellar and sub-stellar subdwarfs \citep{Burgasser2007,
  Kirkpatrick2011}.

Here we discuss two other observational methods that allow the
identification of cool subdwarf candidates among stars in the solar
neighborhood.  These methods utilize the astrometry and photometry
data presented in this paper to evaluate tangential velocities (using
astrometry only), and positions on the H-R diagram (using both
astrometry and photometry) to reveal cool subdwarfs.  After outlining
how subdwarfs can be identified, we apply these two methods on samples
from {\it Gaia} DR1 and the MEarth Project to identify new nearby
subdwarfs.

\subsection{Tangential Velocity}

Kinematic methods that map the motions of stars in our Galaxy have
been used to identify subdwarfs because over billions of years, old
low-metallicity subdwarfs are generally disk-heated to higher spatial
velocities.  The challenge in using $UVW$ kinematics to separate
subdwarfs from dwarfs is that both trigonometric parallaxes {\it and}
radial velocities are required for each candidate star.  Nonetheless,
several efforts have revealed trends.  \cite{Ryan1991} found that
while $\sigma_{U}$, $\sigma_{V}$ and $<$V$>$ are independent of
metallicity, the vertical velocity dispersion, $\sigma_{W}$, increases
with decreasing metallicity.  They use $\sigma_{W}$ values to separate
over 770 FGKM stars from the NLTT catalog, among which they flagged
115 K and 4 M subdwarfs.  A study by \cite{Arifyanto2005}, based on
742 nearby metal-poor stars from \cite{Carney1994}, reported that halo
stars with [Fe/H]$<$$-$1.6 have a low mean rotational velocity around
the Galaxy and a radially elongated velocity ellipsoid, while stars
with $-$1.6$<$[Fe/H]$<$$-$1 have disk-like kinematics.  Based on a
limited sample of 69 subdwarfs with both parallaxes and radial
velocities, \cite{Gizis1997b} found different mean Galactic rotation
velocities between different sub-types of subdwarfs, and noted that
overall, subdwarfs move faster than regular M dwarfs.  A recent result
by \cite{Savcheva2014}, using a much larger sample (3517) drawn from
SDSS data, showed the same trend as \cite{Gizis1997b}.

Without radial velocities, many authors have turned to using a
tangential velocity cutoff to select subdwarfs in lieu of complete
$UVW$ motions.  As a benchmark, \cite{Hawley1996} reported that a
northern sample of 514 M dwarfs has an average tangential velocity of
43.8 km s$^{-1}$.  In order to get an uncontaminated sample of halo
stars from the Giclas high proper motion survey, \cite{Schmidt1975}
imposed a hard limit on tangential velocity of 250 km s$^{-1}$ to
determine the luminosity function of halo stars.  Later,
\cite{Gizis1999} used various tangential velocity cutoffs to select
halo stars from the reduced proper motion diagram in order to revise
the luminosity function of the halo population.  Stars having
$V_{tan}>$ 75 km s$^{-1}$ were flagged as being M extreme subdwarf
candidates, while stars with $V_{tan}>$ 125 km s$^{-1}$ were flagged
as being the M regular subdwarf candidates\footnote{The reduced proper
  motion, ``H'', is defined as
  $H=m+5\log\mu+5=M+5\log\frac{V_{tan}}{4.74}$ where m is the apparent
  magnitude and M is the absolute magnitude.  For a given color, the M
  extreme subdwarfs are fainter than regular subdwarfs. Therefore, M
  extreme subdwarfs do not need to have as high $V_{tan}$ values to be
  placed a few magnitudes below dwarfs on the reduced proper motion
  diagam.}. \cite{Digby2003} used a tangential velocity of 200 km
s$^{-1}$ to select their subdwarf candidates based on the reduced
proper motion diagram.

\begin{figure}
\centering
\includegraphics[angle=-90, scale=0.6]{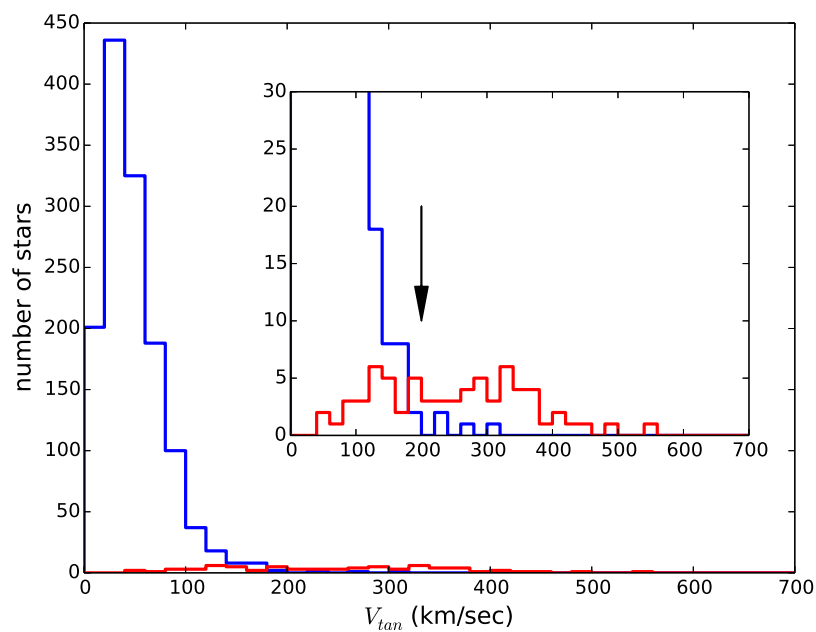}

  \caption{A histogram of tangential velocities for stars in
  different stellar populations.  K and M type stars within 25 pc
  are represented with the blue line, with peak near 30 km
  s$^{-1}$.  Tangential velocities of 167 subdwarfs from our work
  and the literature are shown in red.  The inset panel shows a
  zoomed plot.  The arrow marks the 200 km s$^{-1}$ limit we use to
  select K and M subdwarfs that are found at larger tangential
  velocities.  The four stars with $V_{tan}>$ 200 m s$^{-1}$ shown
  in blue are known subdwarfs within 25 pc.}

\label{fig:vtan}
\end{figure}

Now that we have parallaxes for a relatively large sample of K and M
subdwarfs, we use the sample to determine an appropriate $V_{tan}$
cutoff that can be used to identify subdwarfs.  Figure~\ref{fig:vtan}
illustrates $V_{tan}$ distributions of main sequence dwarfs and
subdwarfs.  The 1324 nearby K and M dwarfs with parallaxes placing
them within 25 pc were selected using spectral types and a color
cutoff of $V-K_s$$>$ 1.9, and are represented with the blue curve.
The red line indicates the 167 K and M subdwarfs observed during our
CTIOPI program with new or revised parallaxes from RECONS,
supplemented with confirmed subdwarfs collected from the literature
\citep{Ryan1991, Carney1994, Hawley1996, Gizis1997b, Cayrel2001,
  Burgasser2003, Lepine2003a, Woolf2005, Jao2009, Wright2014}.  This
plot shows that most K and M dwarfs in the solar neighborhood have
$V_{tan}$ $<$ 100 km s$^{-1}$ with a peak in the distribution at 30 km
s$^{-1}$, consistent with the average tangential velocity discussed in
\cite{Hawley1996} near 45 km s$^{-1}$.  Adopting a cutoff of $V_{tan}$
= 200 km s$^{-1}$ reveals only four stars at higher $V_{tan}$ values,
each of which is, in fact, a known nearby subdwarf.  Thus, there are
no known main sequence K and M dwarfs within 25 pc with $V_{tan}$ $>$
200 km s$^{-1}$.  Unfortunately, this cutoff permits us to identify
only 57\% (95/167) of (fast-moving) subdwarfs and excludes the
remaining 43\% with slower $V_{tan}$ values.  Nonetheless, although
solar neighborhood stars with tangential velocities less than 200 km
s$^{-1}$ comprise a mixed population of young, middle-aged, and old
stars, {\it virtually all nearby stars with tangential velocities
  greater than 200 km s$^{-1}$ are confirmed subdwarfs.}  This cutoff
may be used with confidence to select samples of cool subdwarfs.

\subsection{Hertzsprung-Russell diagram}

\begin{figure}
\centering
\includegraphics[scale=0.8, angle=-90]{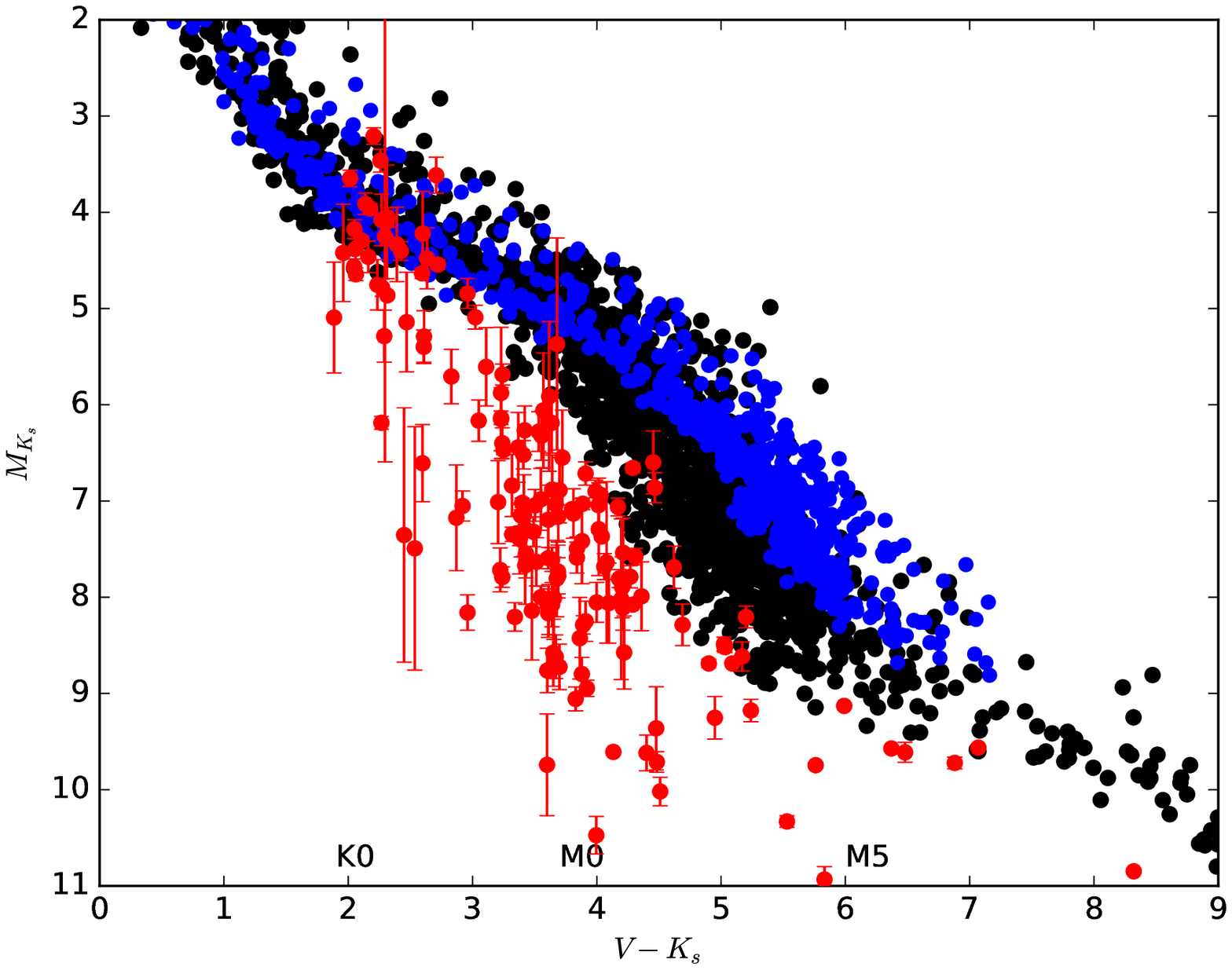}

  \caption{H-R diagram of Pleiades members, nearby stars within 25
  pc, and subdwarfs.  Pleiades (blue points) are from
  \cite{Rebull2016} with a uniform parallax of 7.45$\pm$0.3 mas
  from {\it Gaia} DR1 \citep{GAIA} assigned to all stars.  The
  $V-K_s$ values for the Pleiades members are the dereddened colors
  from \cite{Rebull2016}.  Nearby stars (black points) are the same
  as the gray points in Figure~\ref{fig:HRdiagram} (nearby white
  dwarfs are beyond the boundaries of this plot).  Subdwarfs are
  shown in red.  We omit errorbars for the dwarfs and the Pleiades,
  but their mean absolute magnitude errors are $\pm$0.05 and
  $\pm$0.06, respectively.  Errors in the horizontal direction are
  smaller than the points for all three samples.}

\label{fig:newHR}
\end{figure}

In Figure~\ref{fig:HRdiagram}, we use $M_V$ and $(V-K_S)$ to
illustrate the locations of the sample of objects targeted here.  We
show an additional H-R diagram in Figure~\ref{fig:newHR} using
$V-K_{s}$ vs.~$M_K$, rather than $M_V$.  Stars within 25 pc on the
main sequence are shown in black, overlaid with members of the
Pleiades \citep{Rebull2016} in blue and the 167 subdwarfs discussed in
the previous section in red.  These three sets of stars of various
ages and metallicities are merged blueward of $V-K_{s}\approx$ 2 and
brighter than $M_K\approx$ 4.  In contrast, the three samples form
clear ``bands'' in this diagram in the redder, fainter portion of the
diagram.  Based on Figure~\ref{fig:newHR}, we conclude that: (1) the
general trend shows that the populations' ages, from $\sim$100 Myr for
the Pleiades, to mixed ages of a few Gyr for disk stars, to 6-9 Gyr
for subdwarfs\footnote{\cite{Monteiro2006} measured the ages of two
  cool subdwarfs based on their white dwarf companions.  We adopt this
  age range as representative of the subdwarfs shown here.}), as well
as differing metallicities, causes the shift in the stellar
distribution on the H-R diagram.  (2) Early K-type young stars,
dwarfs, and subdwarfs are indistinguishable in the H-R diagram of
Figure~\ref{fig:newHR}. (3) There is no prominent void in the
distribution of subdwarfs from (at least) $V-K_{S}$ = 2--7 in the
observational H-R diagram.  \cite{Gizis1997b} identified many cool
subdwarfs spectroscopically and established an important foundation in
subdwarf studies.  He found a void on his H-R diagram that lacked
``sdM'' subdwarfs with 2.2$<V-I<$2.8, corresponding to
3.9$<V-K_{s}<$5.2.  Although the number of stars in this region is
still smaller than at bluer colors, the void has begun to fill in
because new subdwarf identification efforts since have increased the
number of subdwarfs in this color range.  Given that the metallicity
distribution is smooth instead of clumped \citep{Gizis1997b}, we
expect more subdwarfs in this color range should be unveiled in the
future.

\section{Subdwarfs Identified using $V_{tan}$ and the H-R diagram}
\label{sec:newsd}

We applied these two methods for identifying new subdwarfs using the
latest large sets of parallaxes recently released in {\it Gaia} DR1
\citep[GAIADR1]{GAIA} which contains stars in common between the
Tycho-2 Catalogue and Gaia mission, and by the MEarth Project which
focuses on nearby M dwarfs \citep{Dittmann2014}.

\subsection{the {\it Gaia} DR1 Catalog}

\begin{figure}
\centering
\includegraphics[scale=0.6, angle=-90.]{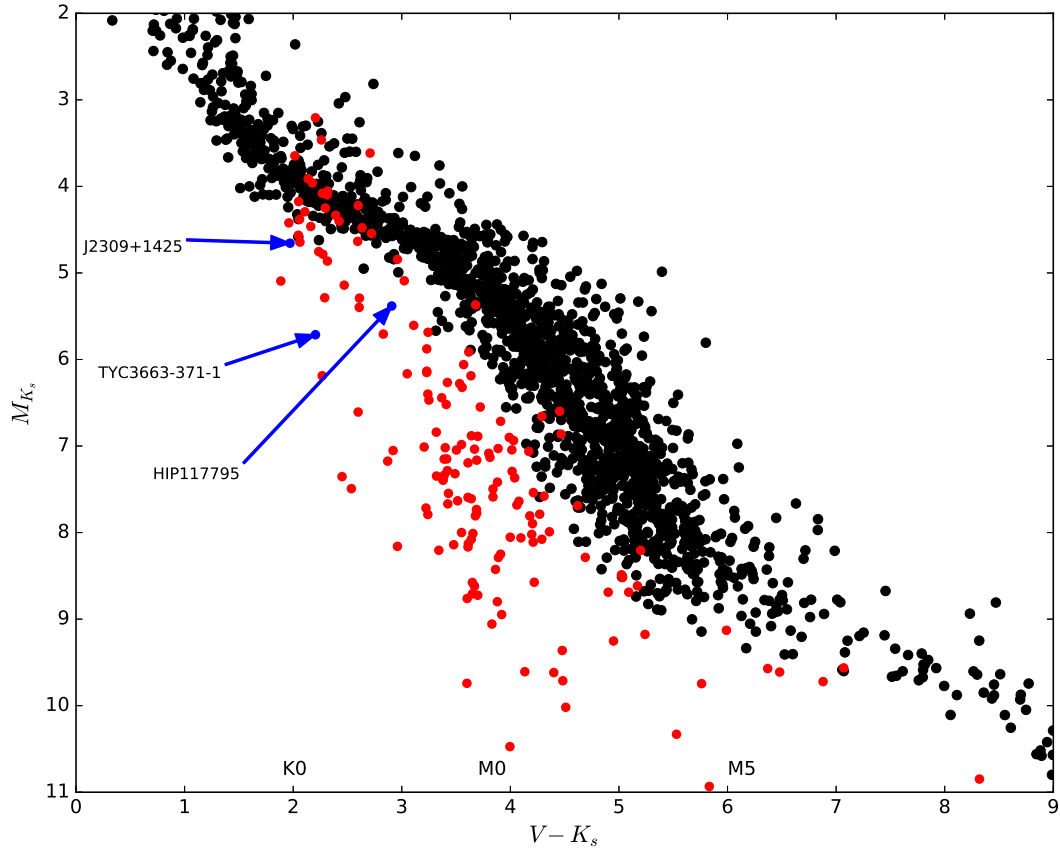}
  \caption{Black points are stars within 25 pc and red points are
  confirmed subdwarfs.  The three labeled blue points are candidate
  subdwarfs within 60 pc below the main sequence from {\it Gaia}
  DR1 and are discussed in Section~\ref{sec:newsd}.  The Johnson $V$
  magnitudes for these three stars are converted from the Tycho 2
  $B$ and $V$ magnitudes.  After further analysis, only TYC
  3663-371-1 remains a subdwarf candidate.}

\label{fig:GAIA}
\end{figure}

We extracted 9494 targets within 60 pc from {\it Gaia} DR1.  Not all
stars have Johnson V magnitudes, so a conversion
($V=V_{T}-0.09*(B_{T}-V_{T})$) from the Tycho 2 catalog was used to
convert Tycho 2 $V$ magnitudes to the standard Johnson $V$ filter.
None of these 9494 targets has a tangential velocity greater than 200
km s$^{-1}$.  We identified only three stars clearly below the main
sequence on the HR diagram.  These three candidates are plotted in
Figure~\ref{fig:GAIA} and discussed below.  We find that two stars are
subdwarf candidates and the third is not a subdwarf. The paucity of
intrinsically faint new subdwarf candidates is not surprising given
that only stars bright enough to be observed by Tycho have parallaxes
in {\it Gaia} DR1.

{\bf (0051+5629) TYC 3663-371-1} This star is almost two magnitudes
fainter than stars in the center of the main-sequence band in
Figure~\ref{fig:GAIA}, providing strong evidence that it is a K
subdwarf.  No metallicity measurement is found in the literature.  

The star is listed as a companion to a G star (HD 4868) in SIMBAD, and
The Washington Double Star Catalog lists the two stars as a visual
binary (WDS 00515+5630AB) with a separation of 40\farcs6
\citep{Hog2000}.  {\it Gaia} DR1 reports a parallax of 21.84$\pm$0.81
for TYC 3663-371-1, but no parallax for HD 4868.  HD 4868 does have a
parallax of 16.28$\pm$0.79 in the {\it Hipparcos} catalog
\citep{vanLeeuwen2007}, which is $\sim$5$\sigma$ offset from TYC
3663-371-1's value.  Both stars have similar proper motions in Tycho
2, but the proper motion for TYC 3663-371-1 in {\it Gaia} DR1 (see
Table~\ref{tbl:TYC3663}) is very different.  Because of the parallax
and proper motion differences, we do not believe these two stars form
a physically bound system.

{\bf (2309+1425) LSPM J2309+1425 = TYC 1167-683-1} There is an X-ray
source in the direction of this star, which is found in a region of
high Galactic latitude molecular clouds \citep{Li2000}.  The ROSAT
catalog shows an X-ray source with $f_{x}/f_{opt}=$$-$1.93 for this
star and \cite{Li2000} reported detection of H$\alpha$
emission. \cite{Cutispoto2002} even set an upper limit on the lithium
equivalent width of 0.8 milli-Angstrom.  The lithium line, X-ray and
H$\alpha$ emission typically indicate youth, so this star is not a
subdwarf. In addition, as shown in Figure~\ref{fig:GAIA}, stars of
different ages merge at this color on the H-R diagram, and this star
is barely offset from main sequence stars.

The star is listed as a companion to HIP 114378 in SIMBAD, and The
Washington Double Star Catalog lists the two stars as a visual binary
(WDS 23100+1426AB) with a separation of 31\farcs7 \citep{Hog2000}.
Both stars have parallaxes and proper motions in {\it Gaia} DR1, given
in Table~\ref{tbl:TYC3663}.  As with TYC 3663-371-1, the distances and
proper motions given in Table~\ref{tbl:TYC3663} do not support that
the two components are physically bound.  Their parallaxes differ by
$\sim$4$\sigma$ and the proper motions in {\it Gaia} DR1 do not match.
Hence, these two stars are likely not associated.

{\bf (2353+5956) HIP 117795} This star has no metallicity measurement
in the literature. \cite{Sperauskas2016} reported a radial velocity of
$-$285.9$\pm$0.2 km s$^{-1}$.  Although this star's tangential
velocity (11.2 km s$^{-1}$) is less than 200 km s$^{-1}$, the {\it
  Gaia} parallax (37.49$\pm$0.23 mas) and proper motion, combined with
the radial velocity measurement, yield $U,V,W$ = (-114, -262, $+$12).
Thus, kinematically, HIP 117795 is very different from nearby disk M
dwarfs \citep{Hawley1996} and it is about one magnitude below the
center of the main sequence in Figure~\ref{fig:GAIA}. As we discussed
earlier, \cite{Arifyanto2005} showed halo stars with [Fe/H]$<$$-$1.6
have low mean galactic rotational velocity, but their velocities range
from $+$200 to $-$200 km s$^{-1}$. As for stars with [Fe/H]$>$$-$1.0
in \cite{Arifyanto2005}, their galactic rotational velocities are
clustered around $+$200 km s$^{-1}$. The direction and velocity of HIP
117795's $V$ indicates it has a retrograde motion and is much faster
than the limit shown in \cite{Arifyanto2005} for halo stars. By
combining its kinematic and location on the H-R diagram, we think it
is likely a nearby K subdwarf.

\subsection{The MEarth Project}

The MEarth Project released 1507 parallaxes of nearby M dwarfs
\citep{Dittmann2014}, but after taking into account companions, there
are a total of 1511 entries.  Although this list of stars lacks
spectral classifications, most of them should be M dwarfs, as MEarth
is targeting M dwarfs.  After calculating each star's tangential
velocity, we find that only LSPM J2107+5943 (LHS 64) has $V_{tan}>
200$ km s$^{-1}$, and it is a known nearby cool subdwarf
\citep{Gizis1997b}.

To reveal additional subdwarfs, we also wish to check the locations of
these stars on the H-R diagram.  \citep{Dittmann2014} provide only
2MASS $J$ and $K_{s}$ magnitudes, and unlike the $V-K_{s}$ color shown
in Figure~\ref{fig:newHR}, $J-K_{s}$ does not clearly differentiate
young M dwarfs, disk M dwarfs, and M subwarfs on the H-R diagram.  In
addition, not all 1511 stars have Johnson $V$ magnitudes, so it is not
possible to overplot the entire sample on the same scale as shown in
Figure~\ref{fig:newHR} for comparison.

In order to create a set of uniform photometry, we cross-matched the
stars with entries in the recent PanSTARRS data release
\citep{Flewelling2016} and extracted $g$ and $r$ magnitudes.  In the
PanSTARRS data release, high proper motion stars usually, but not
always, have multiple entries and coordinates because of the different
epochs of PanSTARRS images, resulting in several lines of matches for
a given coordinate and search radius. Presumably, all of these entries
are the same star moving across the sky. On the other hand, if a star
has only one correct entry in PanSTARRS, we may also get multiple
lines because of several sources within the search radius. In order to
separate correct matches from false ones, we apply multiple steps to
extract PanSTARRS photometry. First, we submitted MEarth stars'
equinox and epoch J2000 coordinates to the PanSTARRS site
(\url{http://archive.stsci.edu/panstarrs/search.php}) and set the
search radius to 0\farcm5.  We calculated the mean epoch from all
entries found in this search. We then slid the MEarth stars'
coordinates from J2000 to that mean epoch using the MEarth proper
motions.  The second search was done by querying these new coordinates
but with a reduced search radius of 0\farcm1, and ``PSFMag'' values
were extracted.  Most stars have only one match in the PanSTARRS
catalog at this mean epoch with the smaller search radius, but for
those stars with multiple matches, a mean photometric magnitude at a
given filter was calculated, with its standard deviation required to
be less than 0.5 mag to eliminate mixing background sources ---
individual photometric measurements for a given filter typically have
errors less than 0.04 mag.  In total, 1349 stars had both $g$ and $r$
magnitudes that were then converted to Johnson $V$ magnitudes using
the equation $V=g-0.5784*(g-r)-0.0038$, which is available at SDSS's
website (\url{
  http://www.sdss3.org/dr8/algorithms/sdssUBVRITransform.php\#West2005}).
We note that PanSTARRS filter bandpasses differ slightly from those of
the SDSS filters.

\begin{figure}
\centering
\includegraphics[scale=0.6, angle=-90.]{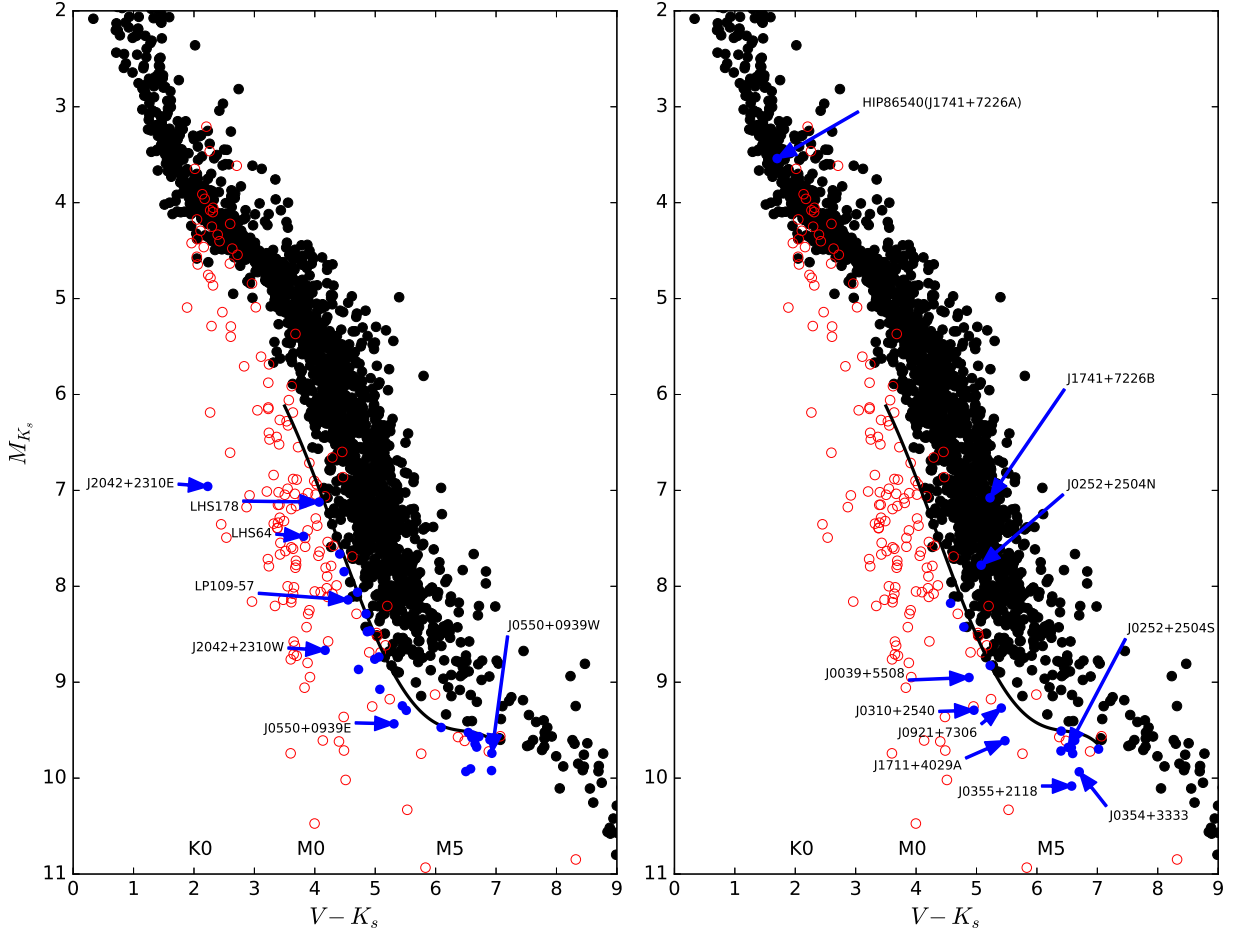}
  \caption{Black points are stars within 25 pc and red open circles
  are confirmed subdwarfs; points are the same in both panels.
  Blue stars in both panels are subdwarf candidates.  The black
  curve is the same in both panels and divides main sequence stars
  from subdwarfs.  The curve is defined by $M_{K_s}=0.019 \times
  color^5-0.42 \times color^4+3.56 \times color^3-14.11 \times
  color^2+27.29 \times color -15.99$, where ``$color$'' is $V-K_s$.
  The left panel highlights stars below the curve that are subdwarf
  candidates.  The right panel highlights stars (generally) below
  the curve that have been spectroscopically classified as main
  sequence M dwarfs.  Labeled stars are discussed in
  section~\ref{sec:newsd}.}

\label{fig:MEarth}
\end{figure}

Rather than relying on our converted $V$ magnitudes, we choose to
identify possible subdwarfs empirically by using cutoffs in $M_{Ks}$
instead of $M_V$ because the 2MASS $K_{s}$ photometry is uniformly
consistent.  The fifth-order polynomal line shown in
Figure~\ref{fig:MEarth} was established based on where known subdwarfs
(represented with red points) are found relative to main sequence
stars (black points).  The 51 MEarth stars located on or below this
dividing line are considered to be subdwarf candidates and are listed
in Table~\ref{tbl:new.sd}.  Three highlighted stars above the lines in
the two panels of Figure~\ref{fig:MEarth} are discussed below.

Among these 51 stars are 30 subdwarf candidates shown with blue points
in the left panel of Figure~\ref{fig:MEarth}.  Blue points in the
right panel represent 20 stars that have been spectroscopically
identified as regular dwarfs in the literature and one (LSPM
J1012+2113) that is a close double with a suspect position in
Figure~\ref{fig:MEarth}.\footnote{LSPM J1012+2113EW was initially
  identified as a subdwarf candidate and a single star, but was
  flagged as having a nearby bright contaminating star by MEarth
  \citep{Newton2016}.  PanSTARRS resolves it as a binary star
  separated by $\sim$2\farcs0 at $\sim$93.3$^{\circ}$ with $\Delta
  g=$0.01.  Without individual $K_{s}$ magnitudes or spectral types,
  we cannot estimate its luminosity class at this time.}.

\subsubsection{New Subwarf Candidates}

Among the 30 subdwarf candidates selected using this method, LHS 64,
LHS 178 and LP 109-57 are previously identified M subdwarfs and are
labeled in the left panel of Figure~\ref{fig:MEarth}.  Thus, we
present here 27 new subdwarf candidates, listed in
Table~\ref{tbl:new.sd}.  We identify two wide subdwarf binary systems,
LSPM J0550+0939EW and LSPM J2042+2310EW, each with both components
falling below the dividing line and labeled in
Figure~\ref{fig:MEarth}.  In particular, LSPM J2042+2310E is $\sim$3
full magnitudes below the main sequence.

\subsubsection{Spectroscopically Confirmed M Dwarfs}

Twenty stars in 17 systems shown in the right panel of
Figure~\ref{fig:MEarth} were previously identified as M dwarfs in the
literature.  Several of these stars are worth discussing to outline
why this group of stars should not be considered subdwarfs.

LSPM J1741+7226B is an M4 dwarf \citep{Alonso2015} located in the
center of main-sequence in Figure~\ref{fig:MEarth}.  It is a wide
common proper motion companion to a bright star, HIP 86540/G 258-16A
\citep{Lepine2007b}.  The MEarth parallax is 77.6$\pm$5.0 mas for LSPM
J1741+7226B, so initially this star was identified as a subdwarf
candidate using the H-R diagram.  However, {\it Gaia} DR1 reports a
parallax of 33.17$\pm$0.21 mas for the primary, so the weighted mean
parallax moves LSPM J1741+7226B upward to the current location shown
in Figure~\ref{fig:MEarth}.  This revised location on the H-R diagram
matches its luminosity, as classified by \cite{Alonso2015}.

Six stars --- LSPM J0039+5508 (LHS 6009), LSPM J0310+2540, LSPM
J0354+3333, LSPM J0355+2118, LSPM J0921+7306 (LHS2126), and LSPM
J1711+4029A (G203-50A)\footnote{This star has an L4 companion
  \citep{Radigan2008} that does not have PanSTARRS photometry so is
  not shown in the plot.} --- highlighted in the right panel of
Figure~\ref{fig:MEarth}, are well below the main sequence, implying
that they are subdwarfs, but their locations contradict their reported
spectral types.  Their 2MASS $K_{s}$ magnitudes should be reliable, so
either their converted $V$ magnitudes or MEarth parallaxes are in
error.  For example, LSPM J0039+5508 has Johnson $V=$14.17 from
\cite{Weis1988}, resulting in $V_{J}-K_{s}=$4.93 compared to
$V_{converted}-K_{s}=$4.88.  The 0.05 magnitude difference is not
sufficient to relocate this star horizontally onto the main sequence,
so if the star is indeed on the main sequence, the parallax is
incorrect.  Thus, we suspect that the parallaxes for these six stars
also need to be revised, as is the case for LSPM J1741+7226.

LSPM J0252+2504N (G36-39) and J0252+2504S form a wide common proper
motion pair.  The primary star is clearly on the main sequence and has
a spectral type of M4.5 \citep{Reid2004}.  Even though the secondary
is below our empirical dividing line, it should be considered a dwarf.
Thus, several of the stars very near the cutoff line are likely just
misplaced in the diagram because of slightly incorrect $V$ estimates
or parallaxes.

\section{The Future of Finding Nearby Subdwarfs}

In the past 20 years, thousands of subdwarfs have been identified
through all-sky spectroscopic surveys or targeted individual
spectroscopic observations.  However, without trigonometric
parallaxes, we cannot pinpoint their locations on the H-R diagram and
link them to their originally defined character: ``dwarfs below the
main sequence'' \citep{Kuiper1939}.  Since Bessel measured the first
parallax in 1838 for 61 Cygni, trigonometric parallaxes continue to be
one of the most essential measurements in stellar astrophysics.  In
this paper we contribute parallaxes for 51 systems, including 37
systems for which these are the first parallaxes, of which 15 have
proper motions of at least 1$\arcsec$ yr$^{-1}$.  We find that most of
the stars targeted here turn out to be cool subdwarfs.

We describe two reliable methods for revealing subdwarfs in the solar
neighborhood.  By using astrometry alone, we find that if a nearby K
or M dwarf has a tangential velocity greater than 200 km s$^{-1}$ it
is almost certainly a subdwarf.  Using accurate trigonometric
parallaxes and $V$ and $K_s$ photometry such as that presented here,
we show that by carefully placing stars on the H-R diagram we can also
identify cool subdwarfs.  In the next a few years, the {\it Gaia} and
LSST efforts will measure countless high precision parallaxes, proper
motions, and photometric values for stars throughout the sky.  We plan
to apply these two methods to identify a large number of nearby low
metallicity subdwarfs, so that we can unveil more of the missing
Galactic relics in the solar neighborhood.

\section{Acknowledgments}

The astrometric observations reported here began as part of the NOAO
Surveys Program in 1999 and continued on the CTIO 0.9-m via the SMARTS
Consortium starting in 2003.  We gratefully acknowledge support from
the National Science Foundation (grants AST 05-07711, AST 09-08402,
and AST 14-12026), NASA's {\it Space Interferometry Mission}, and
Georgia State University, which together have made this long-term
effort possible.  We also thank the members of the SMARTS Consortium,
who enable the operations of the small telescopes at CTIO, as well as
the supporting observers at CTIO, specifically Edgardo Cosgrove,
Arturo G\'{o}mez, Alberto Miranda, and Joselino V\'{a}squez.

The HST-FGS observations were supported under program 11943 by NASA
through grants from the Space Telescope Science Institute, which is
operated by the Association of Universities for Research in Astronomy,
Inc., under NASA contract NAS5-26555.

This research has made use of the SIMBAD database, operated at CDS,
Strasbourg, France.  This publication makes use of data products from
the Two Micron All Sky Survey, which is a joint project of the
University of Massachusetts and the Infrared Processing and Analysis
Center/California Institute of Technology, funded by the National
Aeronautics and Space Administration and the National Science
Foundation.  This work also has made use of data from the European
Space Agency (ESA) mission {\it Gaia}
(\url{https://www.cosmos.esa.int/gaia}), processed by the {\it Gaia}
Data Processing and Analysis Consortium (DPAC,
\url{https://www.cosmos.esa.int/web/gaia/dpac/consortium}).  Funding
for the DPAC has been provided by national institutions, in particular
the institutions participating in the {\it Gaia} Multilateral
Agreement.

The Pan-STARRS1 Surveys (PS1) and the PS1 public science archive,
which were used for this study, have been made possible through
contributions by the Institute for Astronomy, the University of
Hawaii, the Pan-STARRS Project Office, the Max-Planck Society and its
participating institutes, the Max Planck Institute for Astronomy,
Heidelberg, the Max Planck Institute for Extraterrestrial Physics,
Garching, The Johns Hopkins University, Durham University, the
University of Edinburgh, the Queen's University Belfast, the
Harvard-Smithsonian Center for Astrophysics, the Las Cumbres
Observatory Global Telescope Network Incorporated, the National
Central University of Taiwan, the Space Telescope Science Institute,
the National Aeronautics and Space Administration under Grant
No. NNX08AR22G issued through the Planetary Science Division of the
NASA Science Mission Directorate, the National Science Foundation
through Grant No. AST-1238877, the University of Maryland, Eotvos
Lorand University (ELTE), the Los Alamos National Laboratory, and the
Gordon and Betty Moore Foundation.



\clearpage

  
\begin{deluxetable}{lccccccccrrrrrrc}
\rotate
\setlength{\tabcolsep}{0.03in}
\tablewidth{0pt}
\tablecaption{Astrometry Results}
\tabletypesize{\tiny}
\tablehead{\colhead{Name}                &
	   \colhead{RA}                  &
 	   \colhead{DEC}                 &
 	   \colhead{Filt}                &
	   \colhead{Nsea}                &
	   \colhead{Nfrm}                &
	   \colhead{Coverage}            &
	   \colhead{Years}               &
	   \colhead{Nref}                &
	   \colhead{$\pi$(rel)}       &
	   \colhead{$\pi$(corr)}      &
	   \colhead{$\pi$(abs)}       &
	   \colhead{$\mu$}               &
	   \colhead{P.A.}                &
	   \colhead{$V_{tan}$}           &
	   \colhead{Note}                \\
	   \colhead{}                    & 
	   \multicolumn{2}{c}{(J2000.0)} &
           \colhead{}                    &
	   \colhead{}                    &
	   \colhead{}                    &
	   \colhead{}                    &
	   \colhead{}                    &
	   \colhead{}                    &
	   \colhead{(mas)}               &
	   \colhead{(mas)}               &
	   \colhead{(mas)}               &
	   \colhead{(mas/yr)}            &
	   \colhead{(deg)}               &
	   \colhead{(km/s)}              &
	   \colhead{}                    \\
           \colhead{(1)}                 &
           \multicolumn{2}{c}{(2)}       &
           \colhead{(3)}                 &
           \colhead{(4)}                 &
           \colhead{(5)}                 &
           \colhead{(6)}                 &
           \colhead{(7)}                 &
           \colhead{(8)}                 &
           \colhead{(9)}                 &
           \colhead{(10)}                &
           \colhead{(11)}                &
           \colhead{(12)}                &
           \colhead{(13)}                &
           \colhead{(14)}                }
\startdata
\multicolumn{15}{c}{First Trigonometric Parallaxes}\\				   
\hline
LHS 1048        & 00 15 33.51 &  $-$35 11 47.6 & I &  8s &    69 & 2005.71--2012.89 &  7.18 &  8 &   24.47$\pm$1.06 &  1.64$\pm$0.16 &  26.11$\pm$1.07 &  949.4$\pm$0.4 &  100.1$\pm$0.04 &  172.4 &  \\
LHS 127         & 00 55 43.89 &  $-$21 13 07.1 & I &  8s &    61 & 2003.94--2012.94 &  8.99 &  7 &   14.98$\pm$1.10 &  0.52$\pm$0.03 &  15.50$\pm$1.10 & 1227.7$\pm$0.4 &   99.2$\pm$0.03 &  375.6 &  \\
LEHPM 1-1628    & 01 31 05.40 &  $-$50 25 10.0 & I &  7s &    50 & 2005.72--2012.94 &  7.22 &  9 &   10.05$\pm$1.83 &  0.50$\pm$0.06 &  10.55$\pm$1.83 & 1083.0$\pm$0.6 &  142.1$\pm$0.06 &  486.5 &  \\
LHS 1257        & 01 31 30.82 &  $+$10 01 29.7 & I &  7s &    48 & 2005.80--2012.81 &  7.01 &  7 &   20.00$\pm$2.25 &  0.96$\pm$0.13 &  20.96$\pm$2.25 &  929.6$\pm$0.7 &  158.1$\pm$0.08 &  210.2 &  \\
LHS 150         & 02 07 23.26 &  $-$66 34 11.6 & V &  9s &    69 & 2003.95--2012.70 &  8.75 &  8 &   84.91$\pm$1.83 &  1.24$\pm$0.12 &  86.15$\pm$1.83 & 1774.2$\pm$0.7 &   78.2$\pm$0.04 &   97.6 &  \\
LHS 1490        & 03 02 06.36 &  $-$39 50 51.9 & I &  8s &    87 & 2007.55--2015.96 &  8.41 &  9 &   70.46$\pm$1.62 &  0.31$\pm$0.02 &  70.77$\pm$1.62 &  850.6$\pm$0.8 &  220.6$\pm$0.11 &   57.0 &  \\
LHS 1678        & 04 32 42.63 &  $-$39 47 12.3 & V & 11s &   111 & 2003.95--2016.05 & 12.09 &  8 &   49.90$\pm$1.14 &  1.67$\pm$0.14 &  51.57$\pm$1.15 & 1001.0$\pm$0.3 &  166.8$\pm$0.03 &   92.0 &! \\
LEHPM 1-3861    & 05 00 15.78 &  $-$54 06 27.5 & I &  6s &    42 & 2005.90--2011.00 &  5.08 & 10 &   15.58$\pm$1.54 &  0.63$\pm$0.04 &  16.21$\pm$1.54 & 1057.7$\pm$1.1 &  169.2$\pm$0.10 &  309.3 &  \\
LSR 0609+2319   & 06 09 52.43 &  $+$23 19 12.8 & I &  6c &    58 & 2006.05--2011.00 &  4.95 & 11 &   20.88$\pm$1.39 &  1.68$\pm$0.21 &  22.56$\pm$1.41 & 1109.5$\pm$0.8 &  131.3$\pm$0.08 &  233.1 &  \\
SCR 0701-0655   & 07 01 17.79 &  $-$06 55 49.4 & I &  4s &    49 & 2009.08--2011.20 &  3.12 &  9 &    4.68$\pm$0.98 &  1.58$\pm$0.14 &   6.26$\pm$0.99 &  583.3$\pm$0.8 &  185.0$\pm$0.12 &  441.6 &  \\
SCR 0708-4709   & 07 08 32.04 &  $-$47 09 30.6 & V &  5s &    54 & 2007.19--2011.23 &  4.04 &  9 &   11.24$\pm$1.34 &  0.81$\pm$0.10 &  12.05$\pm$1.34 &  402.2$\pm$1.0 &  114.8$\pm$0.28 &  158.3 &  \\
SCR 0709-4648   & 07 09 37.28 &  $-$46 48 58.8 & R &  4c &    53 & 2008.14--2011.23 &  3.09 &  9 &   13.20$\pm$1.41 &  1.22$\pm$0.16 &  14.42$\pm$1.42 &  391.5$\pm$1.2 &    9.0$\pm$0.30 &  128.7 &  \\
SCR 0816-7727   & 08 16 35.65 &  $-$77 27 11.7 & V &  3s &    36 & 2010.01--2012.19 &  2.17 &  9 &   12.72$\pm$1.69 &  0.94$\pm$0.09 &  13.66$\pm$1.69 &  688.3$\pm$2.2 &  324.8$\pm$0.37 &  238.8 &  \\
LHS 2096        & 09 03 08.05 &  $+$08 42 43.8 & R &  4s &    51 & 2010.01--2013.39 &  3.38 &  7 &   16.46$\pm$1.78 &  0.98$\pm$0.14 &  17.44$\pm$1.79 &  549.1$\pm$1.4 &  250.1$\pm$0.27 &  149.3 &  \\
LHS 2099        & 09 05 28.29 &  $-$22 01 56.4 & I &  5s &    48 & 2006.21--2013.38 &  7.18 & 11 &   18.05$\pm$1.07 &  0.55$\pm$0.10 &  18.60$\pm$1.07 &  622.8$\pm$0.8 &  173.3$\pm$0.11 &  158.7 &! \\
LHS 2100        & 09 05 28.29 &  $-$22 01 56.4 & I &  5s &    48 & 2006.21--2013.38 &  7.18 & 11 &   21.33$\pm$1.11 &  0.55$\pm$0.10 &  21.88$\pm$1.11 &  624.0$\pm$0.8 &  173.5$\pm$0.11 &  135.2 &! \\
LHS 2140        & 09 25 31.09 &  $+$00 18 17.6 & I &  6c &    71 & 2008.12--2013.12 &  5.00 & 12 &   16.34$\pm$1.10 &  0.67$\pm$0.04 &  17.01$\pm$1.10 &  577.8$\pm$0.6 &  186.2$\pm$0.10 &  161.0 &! \\
LHS 2299        & 10 42 44.78 &  $-$21 54 20.4 & I &  4c &    41 & 2010.16--2013.26 &  3.10 &  8 &   10.93$\pm$1.40 &  0.93$\pm$0.12 &  11.86$\pm$1.41 &  715.4$\pm$1.1 &  234.2$\pm$0.17 &  285.8 &  \\
SCR 1227-4541   & 12 27 46.83 &  $-$45 41 16.9 & I &  4s &    39 & 2008.07--2011.24 &  3.17 &  9 &   13.37$\pm$1.47 &  1.69$\pm$0.11 &  15.06$\pm$1.47 & 1286.7$\pm$0.8 &  282.5$\pm$0.06 &  404.9 &  \\
SSS 1358-3938   & 13 58 05.40 &  $-$39 37 55.2 & R &  7c &   118 & 2010.16--2016.19 &  6.03 &  9 &   87.06$\pm$0.78 &  1.60$\pm$0.17 &  88.66$\pm$0.80 & 1959.2$\pm$0.5 &  117.2$\pm$0.03 &  104.7 &! \\
LHS 2904        & 14 22 24.92 &  $-$07 17 13.9 & V &  4s &    58 & 2009.32--2012.57 &  3.25 &  9 &   18.61$\pm$2.54 &  1.49$\pm$0.60 &  20.10$\pm$2.61 &  651.5$\pm$2.8 &  247.8$\pm$0.47 &  153.6 &  \\
SCR 1433-3847   & 14 33 03.33 &  $-$38 46 59.6 & I &  4c &    47 & 2008.15--2011.20 &  3.13 &  9 &    7.17$\pm$1.00 &  0.73$\pm$0.09 &   7.90$\pm$1.00 &  471.5$\pm$0.8 &  260.6$\pm$0.17 &  282.8 &  \\
LHS 382         & 14 50 41.22 &  $-$16 56 30.8 & I &  8s &    86 & 2001.21--2011.49 & 10.29 &  7 &   20.69$\pm$0.77 &  0.77$\pm$0.09 &  21.46$\pm$0.78 & 1436.0$\pm$0.2 &  243.8$\pm$0.02 &  317.1 &  \\
SCR 1455-3914   & 14 55 51.56 &  $-$39 14 33.2 & I &  4c &    51 & 2010.17--2013.39 &  3.22 &  8 &   13.79$\pm$0.96 &  0.79$\pm$0.07 &  14.58$\pm$0.96 &  810.8$\pm$1.1 &  266.0$\pm$0.12 &  263.5 &  \\
2MA 1506+1321   & 15 06 54.35 &  $+$13 21 06.1 & I &  7c &    53 & 2010.39--2016.21 &  5.82 & 10 &   86.48$\pm$1.58 &  0.60$\pm$0.04 &  87.08$\pm$1.58 & 1063.0$\pm$0.8 &  270.0$\pm$0.06 &   58.3 &! \\
SIP 1540-2613   & 15 40 29.61 &  $-$26 13 43.0 & I &  3c &    46 & 2010.39--2012.58 &  2.19 & 11 &   66.24$\pm$1.12 &  0.68$\pm$0.12 &  66.92$\pm$1.13 & 1604.4$\pm$1.4 &  225.5$\pm$0.10 &  113.6 &  \\
SCR 1740-5646   & 17 40 46.95 &  $-$56 46 58.1 & I &  5s &    53 & 2008.31--2012.26 &  3.95 &  9 &   13.71$\pm$1.12 &  1.13$\pm$0.08 &  14.84$\pm$1.12 &  447.2$\pm$0.9 &  229.1$\pm$0.23 &  142.9 &  \\
SCR 1756-5927   & 17 56 27.98 &  $-$59 27 18.2 & I &  4s &    38 & 2008.40--2011.70 &  3.33 &  8 &    7.21$\pm$1.38 &  0.75$\pm$0.04 &   7.96$\pm$1.38 &  539.1$\pm$1.0 &  210.8$\pm$0.22 &  321.2 &  \\
SCR 1809-6154B  & 18 09 02.62 &  $-$61 54 14.6 & I &  5s &    55 & 2010.58--2015.39 &  4.81 & 13 &    5.73$\pm$1.64 &  0.49$\pm$0.05 &   6.22$\pm$1.64 &  184.3$\pm$1.2 &  254.4$\pm$0.64 &  140.4 &! \\
SCR 1809-6154A  & 18 09 05.35 &  $-$61 54 14.5 & I &  5s &    27 & 2010.58--2015.39 &  4.81 & 13 &    3.38$\pm$1.64 &  0.49$\pm$0.05 &   3.87$\pm$1.64 &  182.1$\pm$1.2 &  254.3$\pm$0.65 &  222.9 &! \\
G 182-41AB      & 18 09 26.55 &  $+$27 55 23.3 & R &  4c &    52 & 2007.44--2010.65 &  3.22 & 10 &    8.85$\pm$2.33 &  1.11$\pm$0.13 &   9.96$\pm$2.33 &  278.0$\pm$2.1 &  241.5$\pm$0.85 &  132.3 &! \\
WIS 1912-3615   & 19 12 39.24 &  $-$36 14 56.6 & V &  5s &    56 & 2011.50--2015.41 &  3.91 &  9 &   86.22$\pm$1.43 &  1.02$\pm$0.17 &  87.24$\pm$1.44 & 2090.5$\pm$1.0 &  158.2$\pm$0.05 &  113.6 &  \\
SCR 1913-1001   & 19 13 24.63 &  $-$10 01 46.5 & I &  8s &    56 & 2008.70--2015.54 &  6.85 &  8 &    7.90$\pm$0.93 &  2.60$\pm$0.21 &  10.50$\pm$0.95 &  566.3$\pm$0.4 &  211.8$\pm$0.07 &  255.6 &  \\
USN 2101+0307AB & 21 01 04.80 &  $+$03 07 04.7 & I & 10s &    82 & 2006.79--2015.82 &  9.04 &  8 &   55.54$\pm$1.71 &  0.89$\pm$0.10 &  56.43$\pm$1.71 & 1008.0$\pm$0.6 &   91.6$\pm$0.05 &   84.7 &! \\
SCR 2101-5437   & 21 01 45.67 &  $-$54 37 32.0 & I &  4s &    37 & 2008.50--2011.80 &  3.13 & 10 &    9.70$\pm$1.38 &  0.69$\pm$0.03 &  10.39$\pm$1.38 &  711.2$\pm$1.1 &  243.5$\pm$0.16 &  324.5 &  \\
SCR 2204-3347   & 22 04 02.30 &  $-$33 47 38.9 & I &  6s &    49 & 2005.70--2010.75 &  5.03 &  9 &   14.25$\pm$1.36 &  1.51$\pm$0.14 &  15.76$\pm$1.37 &  977.0$\pm$0.7 &  152.4$\pm$0.08 &  293.9 &  \\
LEHPM 1-4592    & 22 21 11.35 &  $-$19 58 14.8 & I &  8c &    60 & 2006.43--2015.56 &  9.13 &  9 &   16.51$\pm$1.11 &  0.33$\pm$0.05 &  16.84$\pm$1.11 & 1059.8$\pm$0.4 &  122.1$\pm$0.04 &  298.4 &  \\
LHS 3841AB      & 22 39 59.41 &  $-$36 15 55.7 & I &  5s &    60 & 2008.70--2012.88 &  4.18 &  7 &   11.73$\pm$1.47 &  0.40$\pm$0.03 &  12.13$\pm$1.47 &  900.2$\pm$1.0 &  170.4$\pm$0.10 &  351.9 &! \\
LHS 539         & 23 15 51.61 &  $-$37 33 30.6 & R &  4s &    57 & 2000.87--2003.77 &  2.89 &  8 &   46.53$\pm$1.00 &  0.92$\pm$0.07 &  47.45$\pm$1.00 & 1309.9$\pm$1.5 &   77.7$\pm$0.11 &  130.9 &  \\
\hline											                     
\multicolumn{15}{c}{Revised Parallaxes}\\				   		                     
\hline											                     
LHS 178         & 03 42 29.45 &  $+$12 31 33.7 & V &  4s &    44 & 2009.93--2012.95 &  3.02 &  8 &   38.37$\pm$2.48 &  1.82$\pm$0.27 &  40.19$\pm$2.49 & 1571.8$\pm$2.2 &  153.4$\pm$0.15 &  185.4 &! \\
G 99-48AB       & 05 59 05.98 &  $+$04 10 38.7 & I &  5c &    55 & 2007.81--2012.88 &  5.07 &  9 &    6.18$\pm$1.83 &  1.88$\pm$0.30 &   8.06$\pm$1.85 &  351.4$\pm$1.5 &  131.6$\pm$0.49 &  206.5 &! \\
LHS 272         & 09 43 46.16 &  $-$17 47 06.2 & V &  9s &    93 & 2001.15--2016.05 & 14.91 & 10 &   68.26$\pm$1.01 &  1.14$\pm$0.11 &  69.40$\pm$1.02 & 1432.5$\pm$0.2 &  279.1$\pm$0.01 &   97.8 &! \\
WT 248          & 10 05 54.94 &  $-$67 21 31.2 & I &  4c &    50 & 2000.14--2003.25 &  3.10 & 11 &   40.52$\pm$2.23 &  1.12$\pm$0.08 &  41.64$\pm$2.23 & 1214.4$\pm$1.8 &  264.8$\pm$0.13 &  138.2 &! \\
G 10-3          & 11 10 02.64 &  $-$02 47 26.4 & V &  7s &    63 & 2010.17--2016.19 &  6.02 &  7 &    6.63$\pm$1.85 &  0.85$\pm$0.17 &   7.48$\pm$1.86 &  493.7$\pm$0.8 &  157.6$\pm$0.18 &  313.1 &! \\
LHS 334         & 12 34 15.78 &  $+$20 37 05.7 & I &  7s &    47 & 2003.24--2011.11 &  7.87 &  6 &   16.78$\pm$1.99 &  0.70$\pm$0.06 &  17.48$\pm$1.99 & 1333.8$\pm$0.7 &  165.6$\pm$0.05 &  361.8 &! \\
LHS 2852        & 14 02 46.66 &  $-$24 31 49.6 & R &  4c &    60 & 2008.20--2011.42 &  3.22 &  8 &   56.43$\pm$1.83 &  1.56$\pm$0.44 &  57.99$\pm$1.88 &  512.8$\pm$1.6 &  317.0$\pm$0.35 &   41.9 &! \\
SSS 1444-2019   & 14 44 20.33 &  $-$20 19 25.5 & I &  6s &    50 & 2010.20--2016.47 &  6.27 &  8 &   59.83$\pm$1.62 &  0.35$\pm$0.03 &  60.18$\pm$1.62 & 3495.1$\pm$1.1 &  235.9$\pm$0.04 &  275.3 &! \\
LHS 385         & 14 55 35.83 &  $-$15 33 44.0 & V &  5s &    49 & 2003.24--2012.58 &  9.34 &  9 &   23.62$\pm$1.51 &  0.81$\pm$0.11 &  24.43$\pm$1.51 & 1727.8$\pm$0.9 &  210.5$\pm$0.06 &  335.2 &! \\
LHS 401         & 15 39 39.06 &  $-$55 09 10.0 & V &  3c &    59 & 2010.16--2012.58 &  2.42 & 10 &   34.40$\pm$1.42 &  4.27$\pm$0.90 &  38.67$\pm$1.68 & 1122.1$\pm$1.8 &  188.2$\pm$0.15 &  137.5 &! \\
LSR 1610-0040AB & 16 10 28.96 &  $-$00 40 54.0 & I & 11s &   140 & 2006.21--2016.19 &  9.98 & 13 &   31.02$\pm$0.46 &  1.24$\pm$0.14 &  32.26$\pm$0.48 & 1448.7$\pm$0.2 &  213.5$\pm$0.01 &  212.8 &! \\
LHS 440AB       & 17 18 25.58 &  $-$43 26 37.6 & R & 12s &   177 & 2000.58--2015.56 & 14.99 & 10 &   37.68$\pm$0.87 &  1.88$\pm$0.54 &  39.56$\pm$1.02 & 1080.2$\pm$0.2 &  233.1$\pm$0.02 &  129.4 &! \\
LHS 456         & 17 50 58.99 &  $-$56 36 06.8 & V &  5s &    49 & 1999.50--2010.50 & 11.10 &  9 &   39.29$\pm$1.43 &  0.58$\pm$0.11 &  39.87$\pm$1.43 & 1256.9$\pm$0.5 &  238.1$\pm$0.04 &  149.4 &! \\
LHS 72          & 23 43 13.65 &  $-$24 09 52.1 & V &  3c &    64 & 2010.50--2012.87 &  2.37 &  7 &   33.23$\pm$1.62 &  1.97$\pm$0.19 &  35.20$\pm$1.63 & 2558.2$\pm$1.9 &  150.2$\pm$0.08 &  344.5 &! \\
LHS 73          & 23 43 13.65 &  $-$24 09 52.1 & V &  3c &    64 & 2010.50--2012.87 &  2.37 &  7 &   30.76$\pm$1.43 &  1.97$\pm$0.19 &  32.73$\pm$1.44 & 2554.6$\pm$1.6 &  150.1$\pm$0.07 &  370.0 &! \\
\enddata

\tablecomments{N$_{sea}$ indicates the number of seasons observed,
  where 4--6 months of observations count as one season, with
  observations typically occurring on 2--3 nights.  The letter ``c''
  indicates a continuous set of observations where multiple nights of
  data were taken in each season, whereas ``s'' indicates scattered
  observations when one or more seasons have only a single night of
  observations.  Generally, ``c'' observations are better.  Stars with
  exclamation marks in the Notes column are discussed in
  Section~\ref{sec:notes}.}

\label{tbl:pi.result}
\end{deluxetable}


\begin{deluxetable}{llrrrcccrrrlc}
\rotate
\tablewidth{0pt}
\tablecaption{Photometry and Spectroscopy Results}
\tabletypesize{\tiny}
\tablehead{
           \colhead{Name1}   &
           \colhead{Name2}   &
           \colhead{$V$}     &
           \colhead{$R$}     &
           \colhead{$I$}     &
           \colhead{\#}      &
	   \colhead{$\pi$}   &
	   \colhead{$\sigma$}&
           \colhead{$J$}     &
           \colhead{$H$}     &
           \colhead{$K_{s}$} &
	   \colhead{Spect.}  &
	   \colhead{Refs}    \\
	   \colhead{}        &
	   \colhead{}        &
	   \colhead{mag}     &
	   \colhead{mag}     &
	   \colhead{mag}     &
	   \colhead{}        &
	   \colhead{filter}  &
	   \colhead{mag}     &
	   \colhead{mag}     &
	   \colhead{mag}     &
	   \colhead{mag}     &
	   \colhead{}        &
	   \colhead{}        \\
	   \colhead{(1)}     &
	   \colhead{(2)}     &
	   \colhead{(3)}     &
	   \colhead{(4)}     &
	   \colhead{(5)}     &
	   \colhead{(6)}     &
	   \colhead{(7)}     &
           \colhead{(8)}     &
           \colhead{(9)}     &
           \colhead{(10)}    &
	   \colhead{(11)}    &
	   \colhead{(12)}    & 
	   \colhead{(13)}     
           }
\startdata
LHS 1048        & G 267-58         & 14.53 & 13.47 & 12.09 &  2& I & 0.0094 & 10.80$\pm$0.02 & 10.26$\pm$0.02 & 10.06$\pm$0.02 &  M4       &  1 \\
LHS 127         & G 268-77         & 15.79 & 14.77 & 13.61 &  2& I & 0.0091 & 12.46$\pm$0.02 & 11.92$\pm$0.02 & 11.73$\pm$0.02 &  M2.0VI   &  8 \\
LEHPM 1-1628    &                  & 17.11 & 16.16 & 15.23 &  2& I & 0.0197 & 14.14$\pm$0.03 & 13.71$\pm$0.03 & 13.46$\pm$0.04 &  M1.0VI   &  8 \\
LHS 1257        & LSPM J0131+1001  & 16.37 & 15.26 & 13.82 &  2& I & 0.0103 & 12.40$\pm$0.02 & 11.93$\pm$0.02 & 11.68$\pm$0.02 &  VI       & 16 \\
LHS 150         & GJ 85            & 11.49 & 10.49 &  9.31 &  3& V & 0.0092 &  8.13$\pm$0.02 &  7.61$\pm$0.03 &  7.36$\pm$0.02 &  M1.5V    &  6 \\
LHS 1490        & LP 994-33        & 15.87 & 14.35 & 12.44 &  3& I & 0.0104 & 10.71$\pm$0.02 & 10.18$\pm$0.03 &  9.88$\pm$0.02 &  M5.0VI   &  8 \\
LHS 1678        & LP 375-2         & 12.48 & 11.46 & 10.26 &  3& V & 0.0064 &  9.02$\pm$0.03 &  8.50$\pm$0.05 &  8.26$\pm$0.03 &  M2.0V    & 12 \\
LEHPM 1-3861    & SSSPM J0500-5406 & 18.48 & 17.24 & 15.77 &  2& I & 0.0107 & 14.44$\pm$0.03 & 14.12$\pm$0.05 & 13.97$\pm$0.06 &  M4.0VI   &  8 \\
LSR 0609+2319   & LSPM J0609+2319  & 17.65 & 16.33 & 14.65 &  2& I & 0.0078 & 13.16$\pm$0.02 & 12.64$\pm$0.02 & 12.41$\pm$0.02 &  sdM5.0   & 10 \\
SCR 0701-0655   &                  & 16.55 & 15.61 & 14.75 &  2& I & 0.0088 & 13.73$\pm$0.02 & 13.19$\pm$0.02 & 13.00$\pm$0.03 &  M1.0VI   &  8 \\
SCR 0708-4709   &                  & 13.81 & 13.05 & 12.37 &  3& V & 0.0078 & 11.44$\pm$0.02 & 10.90$\pm$0.02 & 10.76$\pm$0.03 &  K7.0VI   &  8 \\
SCR 0709-4648   & PM J07096-4648   & 14.91 & 14.02 & 13.22 &  2& R & 0.0088 & 12.20$\pm$0.03 & 11.70$\pm$0.03 & 11.49$\pm$0.03 &  M0.5VI   &  8 \\
SCR 0816-7727   &                  & 15.30 & 14.40 & 13.58 &  3& V & 0.0070 & 12.62$\pm$0.03 & 12.07$\pm$0.02 & 11.87$\pm$0.02 &  VI       & 16 \\
LHS 2096        & LP 486-42        & 17.81 & 16.64 & 15.29 &  3& R & 0.0093 & 13.99$\pm$0.02 & 13.58$\pm$0.03 & 13.41$\pm$0.04 & esdM5.5   & 11 \\
LHS 2099        & LP 845-16        & 15.83 & 14.84 & 13.79 &  3& I & 0.0088 & 12.64$\pm$0.03 & 12.15$\pm$0.04 & 11.94$\pm$0.03 & esdM2.0   & 11 \\
LHS 2100        & LP 845-17        & 19.16 & 17.73 & 15.80 &  3& I & 0.0170 & 14.27$\pm$0.03 & 13.87$\pm$0.04 & 13.63$\pm$0.05 & esdM2.0   & 11 \\
LHS 2140        & G 46-40          & 15.05 & 14.13 & 13.24 &  4& I & 0.0099 & 12.17$\pm$0.02 & 11.63$\pm$0.02 & 11.44$\pm$0.03 &  VI       & 3  \\
LHS 2139        &                  & 19.56 & 18.93 & 18.34 &  4& I &\nodata & \nodata        &  \nodata       &  \nodata       & WD        & 3  \\
LHS 2299        & LP 790-36        & 16.95 & 15.90 & 14.69 &  2& I & 0.0077 & 13.52$\pm$0.02 & 12.98$\pm$0.02 & 12.74$\pm$0.03 &  sdM3.0   & 11 \\
SCR 1227-4541   & PM J12277-4541   & 15.23 & 14.40 & 13.69 &  2& I & 0.0079 & 12.75$\pm$0.03 & 12.39$\pm$0.03 & 12.27$\pm$0.03 &  VI       & 16 \\
SSS 1358-3938   &                  & 14.04 & 12.80 & 11.20 &  2& R & 0.0110 &  9.72$\pm$0.02 &  9.23$\pm$0.02 &  8.95$\pm$0.02 &  VI       & 16 \\
LHS 2904        & G 124-29         & 12.40 & 11.60 & 10.87 &  3& V & 0.0074 &  9.93$\pm$0.02 &  9.31$\pm$0.02 &  9.15$\pm$0.02 &  VI       & 16 \\
SCR 1433-3847   & PM J14330-3846   & 17.23 & 16.29 & 15.41 &  3& I & 0.0086 & 14.37$\pm$0.04 & 13.78$\pm$0.05 & 13.59$\pm$0.05 &  M0.5VI   &  8 \\
LHS 382         & LP 801-16        & 15.73 & 14.61 & 13.16 &  2& I & 0.0079 & 11.85$\pm$0.02 & 11.38$\pm$0.03 & 11.11$\pm$0.02 &  M1.5     & 15 \\
SCR 1455-3914   & PM J14558-3914   & 15.43 & 14.50 & 13.58 &  2& I & 0.0087 & 12.50$\pm$0.02 & 11.98$\pm$0.02 & 11.79$\pm$0.02 &  M1.0VI   &  8 \\
2MA 1506+1321   &                  &\nodata& 19.30 & 16.93 &  2& I & 0.0286 & 13.37$\pm$0.02 & 12.38$\pm$0.02 & 11.74$\pm$0.02 &  L3.0     &  4 \\
SIP 1540-2613   &                  & 19.21 & 16.57 & 14.11 &  1& I & 0.0089 & 11.65$\pm$0.03 & 11.14$\pm$0.03 & 10.73$\pm$0.02 &  V        & 16 \\
SCR 1740-5646   &                  & 17.03 & 16.01 & 14.95 &  3& I & 0.0099 & 13.83$\pm$0.03 & 13.33$\pm$0.03 & 13.20$\pm$0.04 &  M3.0VI   &  8 \\
SCR 1756-5927   &                  & 16.30 & 15.38 & 14.49 &  2& I & 0.0092 & 13.44$\pm$0.03 & 12.89$\pm$0.03 & 12.69$\pm$0.03 &  M1.0VI   &  8 \\
SCR 1809-6154A  &                  & 15.66 & 14.79 & 13.92 &  2& I &\nodata & 12.84$\pm$0.03 & 12.32$\pm$0.03 & 12.09$\pm$0.02 &  VI       & 16 \\
SCR 1809-6154B  &                  & 16.11 & 15.19 & 14.21 &  2& I & 0.0094 & 13.14$\pm$0.03 & 12.57$\pm$0.03 & 12.43$\pm$0.02 &  VI       & 16 \\
G 182-41AB      & LP 334-10        & 12.62J& 12.06J& 11.53J&  3& R & 0.0088 & 10.74$\pm$0.02J& 10.26$\pm$0.03J& 10.15$\pm$0.02 &  VI       & 16 \\
WIS 1912-3615   &                  & 13.91 & 12.64 & 11.00 &  3& V & 0.0095 &  9.52$\pm$0.02 &  9.01$\pm$0.06 &  8.77$\pm$0.02 &  mid-M    &  5 \\
SCR 1913-1001   &                  & 15.60 & 14.66 & 13.80 &  3& I & 0.0119 & 12.71$\pm$0.03 & 12.16$\pm$0.03 & 11.93$\pm$0.03 &  VI       & 16 \\
USN 2101+0307AB &                  & 18.67J& 16.64J& 14.32J&  3& I & 0.0090 & 11.70$\pm$0.02J& 10.96$\pm$0.02J& 10.57$\pm$0.02J&  V        & 16 \\
SCR 2101-5437   &                  & 15.77 & 14.84 & 13.86 &  3& I & 0.0066 & 12.79$\pm$0.03 & 12.26$\pm$0.02 & 12.08$\pm$0.03 &  M1.0VI   &  8 \\
SCR 2204-3347   &                  & 15.44 & 14.45 & 13.41 &  3& I & 0.0074 & 12.32$\pm$0.03 & 11.81$\pm$0.03 & 11.60$\pm$0.03 &  M3.0VI   &  8 \\
LEHPM 1-4592    &                  & 19.96 & 18.16 & 15.98 &  3& I & 0.0091 & 14.19$\pm$0.03 & 13.74$\pm$0.04 & 13.48$\pm$0.04 &  VI       & 16 \\
LHS 3841AB      & LP 984-76        & 16.87J& 15.89J& 14.89J&  2& I & 0.0061 & 13.82$\pm$0.02J& 13.32$\pm$0.03J& 13.20$\pm$0.04J&  sdM2.5   & 11 \\
LHS 539         & LP 986-16        & 14.97 & 13.66 & 11.98 &  3& R & 0.0107 & 10.40$\pm$0.02 &  9.87$\pm$0.02 &  9.59$\pm$0.02 &  V        & 16 \\
\hline										     												       
LHS 178         & G 79-59          & 12.87 & 11.89 & 10.78 &  2& V & 0.0076 &  9.60$\pm$0.02 &  9.11$\pm$0.02 &  8.88$\pm$0.02 &  sdM1.5   &  3 \\
G 99-48AB       & LTT 17896AB      & 11.85J& 11.42J& 10.97J&  3& I & 0.0086 & 10.38$\pm$0.03J&  9.99$\pm$0.02J&  9.89$\pm$0.02J&  VI       & 16 \\
LHS 272         & LP 788-27        & 13.16 & 12.10 & 10.87 &  3& V & 0.0125 &  9.62$\pm$0.02 &  9.12$\pm$0.02 &  8.87$\pm$0.02 &  M3.0VI   &  9 \\
WT 248          &                  & 14.52 & 13.40 & 11.95 &  2& I & 0.0077 & 10.56$\pm$0.02 & 10.10$\pm$0.02 &  9.87$\pm$0.02 &  M3.0V    &  7 \\ 
G 10-3          & LHS 2361         & 12.56 & 12.01 & 11.50 &  2& V & 0.0107 & 10.73$\pm$0.02 & 10.27$\pm$0.02 & 10.11$\pm$0.02 &  VI       &  8 \\
LHS 334         & LP 377-13        & 17.99 & 16.75 & 15.13 &  3& I & 0.0102 & 13.75$\pm$0.03 & 13.25$\pm$0.04 & 13.04$\pm$0.03 &  M6.0VI   &  8 \\
LHS 2852        & LP 856-36        & 12.13 & 11.08 &  9.85 &  2& R & 0.0242 &  8.63$\pm$0.03 &  8.10$\pm$0.03 &  7.84$\pm$0.02 &  sdM2.0   &  3 \\
SSS 1444-2019   & LP 741-20        & 20.25 & 17.62 & 14.95 &  4& I & 0.0130 & 12.55$\pm$0.03 & 12.14$\pm$0.03 & 11.93$\pm$0.03 &  sdM9     & 14 \\
LHS 385         & LP 741-20        & 14.61 & 13.67 & 12.78 &  3& V & 0.0087 & 11.74$\pm$0.03 & 11.28$\pm$0.02 & 11.06$\pm$0.02 &  M1.0VI   &  8 \\
LHS 401         & L 201-12         & 12.73 & 11.88 & 11.12 &  2& V & 0.0077 & 10.15$\pm$0.02 &  9.60$\pm$0.02 &  9.41$\pm$0.02 &  M0.5VI   &  8 \\
LSR 1610-0040AB &                  & 19.09J& 17.10J& 14.97J&  2& I & 0.0077 & 12.91$\pm$0.02J& 12.30$\pm$0.02J& 12.02$\pm$0.03J&  sd?M6pec &  2 \\
LHS 440AB       & L 413-156        & 12.98J& 11.98J& 10.87J&  3& R & 0.0113 &  9.70$\pm$0.02J&  9.13$\pm$0.02J&  8.95$\pm$0.02J&  M1.0VI   &  8 \\
LHS 456         & L 205-83         & 12.08 & 11.12 & 10.09 &  3& V & 0.1282 &  8.99$\pm$0.02 &  8.42$\pm$0.02 &  8.19$\pm$0.02 &  M2.0     & 15 \\
LHS 72          & G 275-90         & 12.07 & 11.24 & 10.49 &  3& V & 0.0087 &  9.61$\pm$0.03 &  9.04$\pm$0.02 &  8.82$\pm$0.02 &  VI       & 13 \\
LHS 73          & G 275-92         & 12.77 & 11.90 & 11.10 &  3& V & 0.0093 & 10.11$\pm$0.02 &  9.59$\pm$0.02 &  9.37$\pm$0.02 &  K6.0VI   &  8 \\

\enddata

\tablecomments{Stars without spectral types are noted with luminosity
  class only in column 12, based on their locations on the H-R diagram
  or independent metallicity measurements.  All of these stars are
  labeled in Figure~\ref{fig:HRdiagram}. ``J'' next to a magnitude
  indicates a combined photometry.}

\tablerefs{
 (1) \citealt{Bidelman1985};
 (2) \citealt{Dahn2008};  
 (3) \citealt{Gizis1997}; 
 (4) \citealt{Gizis2000};
 (5) \citealt{Gizis2011}; 
 (6) \citealt{Hawley1996};
 (7) \citealt{Henry2002}; 
 (8) \citealt{Jao2008};   
 (9) \citealt{Jao2011};   
(10) \citealt{Reid2003};  
(11) \citealt{Reid2005};  
(12) \citealt{Reid2007};  
(13) \citealt{Rodgers1974}
(14) \citealt{Scholz2004};
(15) \citealt{Walker1983};
(16) this work
}

\label{tbl:phot.result}
\end{deluxetable}

\begin{deluxetable}{lclcccc}
\rotate
\tablewidth{0pt}
\tablecaption{Two Wide Binaries Selected from {\it Gaia} DR1}
\tabletypesize{\small}
\tablehead{\colhead{Name}   &
           \colhead{$\pi$}   &
           \colhead{Ref}  &
           \colhead{Tycho2 $\mu_{RA}$}     &
           \colhead{Tycho2 $\mu_{Decl.}$}  &
           \colhead{{\it Gaia} DR1 $\mu_{RA}$}     &
           \colhead{{\it Gaia} DR1 $\mu_{Decl.}$}
           \\
           \colhead{} &
           \colhead{mas} &
           \colhead{} &
           \colhead{mas} &
           \colhead{mas} &
           \colhead{mas} &
           \colhead{mas} 
           }
\startdata
HD 4868\tablenotemark{a}        &  16.28$\pm$0.79 & {\it Hipparcos} &   62.2$\pm$1.5 & -80.9$\pm$1.5 &  \nodata         & \nodata         \\ 
TYC 3663-371-1\tablenotemark{a} &  21.84$\pm$0.81 & {\it Gaia} DR1  &   69.2$\pm$5.8 & -88.7$\pm$5.8 &  46.43$\pm$1.84  & -66.98$\pm$1.57 \\
\hline
HIP 114378\tablenotemark{b}     &  39.87$\pm$0.41 & {\it Gaia} DR1  & -121.7$\pm$4   & -84.8$\pm$5   & -121.55$\pm$0.04 & -85.36$\pm$0.03 \\
TYC 1167-683-1\tablenotemark{b} &  36.55$\pm$0.75 & {\it Gaia} DR1  & -125.8$\pm$13  & -96.3$\pm$14  & -107.25$\pm$0.80 & -91.25$\pm$0.86 \\
\enddata

\label{tbl:TYC3663}

\tablenotetext{a,b} {We find that neither of these two systems form binary systems.}

\end{deluxetable}

\begin{deluxetable}{lllllllcclc}
\rotate
\tablewidth{0pt}
\tablecaption{Subdwarf Candidates Selected from MEarth}
\tabletypesize{\small}
\tablehead{\colhead{R.A.}   &
           \colhead{Decl.}  &
           \colhead{Name1}  &
           \colhead{Name2}  &
           \colhead{$g$}    &
           \colhead{$r$}    &
           \colhead{$V$}    &
           \colhead{$V-K_{s}$} &
           \colhead{$M_{K_s}$} &
           \colhead{SpT}  &
           \colhead{Ref}}

\startdata
\multicolumn{11}{c}{Previously Identified as Subdwarfs}\\
\hline
03 42 30.06&+12 31 16.20&LSPM J0342+1231  &LHS 178   &13.55&12.52&12.95& 4.07& 7.12& sdM1.5 & 1 \\
21 07 53.68&+59 42 56.00&LSPM J2107+5943  &LHS 64    &13.92&12.69&13.20& 3.81& 7.48& sdM1.5 & 1 \\
22 58 15.55&+61 44 26.30&LSPM J2258+6144  &LP 109-57 &14.78&13.46&14.01& 4.55& 8.14& sdM3   & 5 \\
\hline
\multicolumn{11}{c}{New Subdwarf Candidates}\\
\hline
00 34 37.89&+40 49 59.50&LSPM J0034+4050  &          &18.17&16.81&17.38& 6.50& 9.93&        & \\
00 46 35.83&+36 36 36.90&LSPM J0046+3636  &G132-28   &14.82&13.63&14.12& 4.48& 7.85&        & \\
01 41 54.90&+38 43 25.00&LSPM J0141+3843  &          &15.60&14.32&14.85& 4.91& 8.47&        & \\
01 43 53.38&+00 14 31.10&LSPM J0143+0014  &          &17.93&16.60&17.16& 6.90& 9.60&        & \\
03 05 35.80&+19 34 06.80&LSPM J0305+1934  &          &15.57&14.37&14.87& 4.99& 8.76&        & \\
03 14 12.77&+28 40 30.10&LSPM J0314+2840  &LHS 1516  &17.47&16.12&16.69& 6.60& 9.58&        & \\
03 36 22.62&+13 50 38.90&LSPM J0336+1350  &LHS 1568  &15.52&14.37&14.85& 4.85& 8.29&        & \\
04 19 25.55&+38 15 01.50&LSPM J0419+3815  &          &14.80&13.54&14.07& 4.71& 8.06&        & \\
04 28 49.51&+07 28 29.40&LSPM J0428+0728  &LHS 5097  &18.82&17.63&18.13& 6.93& 9.92&        & \\
04 52 29.45&+09 30 24.60&LSPM J0452+0930  &          &16.74&15.41&15.96& 5.51& 9.29&        & \\
05 42 29.77&+07 31 05.30&LSPM J0542+0731  &G 102-24  &15.22&13.98&14.50& 4.87& 8.47&        & \\
05 50 11.21&+09 40 03.70&LSPM J0550+0939E &          &16.41&15.09&15.64& 5.31& 9.43&        & \\
05 50 11.21&+09 40 03.70&LSPM J0550+0940W &          &18.38&16.99&17.57& 6.93& 9.74&        & \\
06 37 55.42&+08 58 55.10&LSPM J0637+0858  &          &16.45&15.18&15.71& 5.45& 9.25&        & \\
07 31 29.26&+02 49 08.90&LSPM J0731+0249  &          &17.63&16.30&16.85& 6.61& 9.55&        & \\
08 01 21.10&+56 24 00.40&LSPM J0801+5624  &          &18.28&16.91&17.48& 6.65& 9.64&        & \\
09 17 07.19&+20 07 51.30&LSPM J0917+2007  &G 41-30   &15.32&14.19&14.66& 5.06& 8.74&        & \\
09 19 20.07&+21 54 28.40&LSPM J0919+2154  &LP 369-27 &18.86&17.49&18.07& 6.72& 9.57&        & \\
11 59 58.94&+21 04 59.90&LSPM J1159+2105  &LP 375-69 &17.46&16.20&16.73& 6.09& 9.47&        & \\
16 51 05.15&+78 09 23.10&LSPM J1651+7809  &LHS 3247  &17.77&16.45&17.00& 6.54& 9.52&        & \\
17 47 26.14&+28 40 38.10&LSPM J1747+2840  &LHS 6325  &15.57&14.28&14.82& 5.08& 9.07&        & \\
17 57 00.01&+78 59 50.90&LSPM J1756+7859  &          &18.71&17.40&17.95& 6.58& 9.90&        & \\
19 12 45.12&+39 43 20.20&LSPM J1912+3943  &          &15.29&14.08&14.58& 4.72& 8.87&        & \\
20 32 09.80&+60 18 16.80&LSPM J2032+6018  &          &15.04&13.79&14.31& 4.41& 7.66&        & \\
20 42 29.08&+23 10 13.50&LSPM J2042+2310E &          &11.20&10.60&10.85& 2.23& 6.96&        & \\
20 42 29.12&+23 10 14.90&LSPM J2042+2310W &          &15.21&13.99&14.50& 4.17& 8.67&        & \\
23 29 25.36&+46 26 38.70&LSPM J2329+4626  &          &18.68&17.36&17.91& 6.68& 9.68&        & \\
\hline
\multicolumn{11}{c}{Previously Identified as Main Sequence Dwarfs}\\
\hline
00 39 18.36&+55 08 10.10&LSPM J0039+5508  & LHS 6009 &14.84&13.60&14.12& 4.88& 8.95& M3.5   & 1 \\
00 43 35.48&+28 26 28.40&LSPM J0043+2826  & LHS 120  &15.18&13.95&14.46& 4.79& 8.42& M4     & 2 \\
01 11 36.69&+41 27 51.70&LSPM J0111+4127  &LP 194-35 &19.05&17.65&18.24& 7.02& 9.70& M5.5   & 3 \\
02 08 14.23&+49 48 55.10&LSPM J0208+4949  & LHS 1345 &18.44&17.18&17.71& 6.57& 9.68& M5.5   & 4 \\
02 52 33.27&+25 04 46.70&LSPM J0252+2504N &G 36-39A  &15.48&14.30&14.80& 5.08& 7.78& M4.5   & 5 \\
02 52 34.18&+25 04 33.10&LSPM J0252+2504S &G 36-39B  &18.90&17.59&18.14& 6.52& 9.68&\nodata &   \\
03 10 38.98&+25 40 51.00&LSPM J0310+2540  &LP 355-32 &14.85&13.62&14.14& 4.96& 9.29& M3     & 6 \\
03 54 01.36&+33 33 21.40&LSPM J0354+3333  &          &18.39&17.07&17.62& 6.70& 9.93& M6     & 7 \\
03 55 37.16&+21 18 47.60&LSPM J0355+2118  &LP 357-206&18.44&17.13&17.67& 6.57&10.08& M5     & 8 \\
07 29 18.83&+75 53 58.80&LSPM J0729+7554  &LP 17 44  &18.45&17.20&17.72& 6.59& 9.74& M5.5   & 5 \\
09 20 22.75&+26 43 36.20&LSPM J0920+2643  &LHS 266   &16.30&14.95&15.52& 5.23& 8.83& M4.5   & 2 \\
09 21 16.79&+73 06 34.20&LSPM J0921+7306  &LHS 2126  &15.72&14.36&14.93& 5.41& 9.27& M4.5   & 2 \\
13 57 00.62&+08 30 09.80&LSPM J1357+0830  &LHS 2828  &18.31&16.99&17.54& 6.40& 9.51& M5.5   & 4 \\
16 37 01.25&+35 35 38.40&LSPM J1637+3535  &LHS 3227  &17.41&16.08&16.64& 6.40& 9.72& M6     & 6 \\
17 11 46.38&+40 29 02.60&LSPM J1711+4029A &G 203-50A &16.50&15.19&15.74& 5.47& 9.61& M4.5   & 9 \\
17 11 46.38&+40 29 02.60&LSPM J1711+4029B &G 203-50B &\nodata&\nodata&\nodata&\nodata&\nodata& L4 & 9 \\
17 41 06.69&+72 25 13.24&LSPM J1741+7225A &G 258-16  &\nodata&\nodata&7.61&1.70& 3.54&K0    & 10 \\
17 41 15.76&+72 26 34.80&LSPM J1741+7226B &G 258-17  &15.39&14.14&14.66& 5.22& 7.08& M4.0   & 11 \\
18 41 47.82&+24 21 50.80&LSPM J1841+2421  &          &18.16&16.83&17.39& 6.63& 9.61& M6.0   & 7 \\
22 56 14.09&+68 15 32.70&LSPM J2256+6815  &LHS 3877  &15.33&14.11&14.62& 4.57& 8.18& M3.5   & 2 \\
\hline
\multicolumn{11}{c}{Binary, Presumed Main Sequence Dwarfs}\\
\hline
10 12 58.30&+21 13 22.20&LSPM J1012+2113EW&          &99.13&15.86&50.97&39.82& 8.91&\nodata & New Double  \\
\hline
\enddata
\tablerefs{ 
(1) \citealt{Gizis1997}
(2) \citealt{Hawley1996}
(3) \citealt{Cruz2002}
(4) \citealt{Reid2005}
(5) \citealt{Reid2004}
(6) \citealt{Scholz2005}
(7) \citealt{Lepine2003b}
(8) \citealt{Cruz2002}
(9) \citealt{Radigan2008}
(10) \citealt{White2007}
(11) \citealt{Alonso2015}
}
\label{tbl:new.sd}
\end{deluxetable}

\end{document}